\documentclass[manuscript]{aastex63}

\usepackage{amsmath}
\usepackage[caption=false]{subfig}
\usepackage{float}

\received{2021 May 07}
\revised{2021 June 10}
\accepted{2021 June 25}

\submitjournal{Publications of the Astronomical Society of the Pacific}

\shorttitle{Asteroid Photometry with PIRATE}
\shortauthors{S.~L. Jackson, et al.}

\begin{document}

\title{Asteroid Photometry with PIRATE: \\Optimizations and Techniques for Small Aperture Telescopes}

\correspondingauthor{Samuel L. Jackson}
\email{samuel.jackson@open.ac.uk}

\author[0000-0001-9242-4254]{Samuel L. Jackson}
\affiliation{School of Physical Sciences, The Open University, \\
Milton Keynes, MK7 6AA, UK}

\author[0000-0001-8670-8365]{Ulrich C. Kolb}
\affiliation{School of Physical Sciences, The Open University, \\
Milton Keynes, MK7 6AA, UK}

\author[0000-0002-9153-9786]{Simon F. Green}
\affiliation{School of Physical Sciences, The Open University, \\
Milton Keynes, MK7 6AA, UK}

\begin{abstract}

    Small aperture telescopes provide the opportunity to conduct high frequency, targeted observations of near-Earth Asteroids that are not feasible with larger facilities due to highly competitive time allocation requirements.
    Observations of asteroids with these types of facilities often focus on rotational brightness variations rather than longer-term phase angle dependent variations (phase curves) due to the difficulty of achieving high precision photometric calibration.
    We have developed an automated asteroid light curve extraction and calibration pipeline for images of moving objects from the $0.43\,$m Physics Innovations Robotic Telescope Explorer (PIRATE).
    This allows for the frequency and quality of observations required to construct asteroid phase curves.
    Optimisations in standard data reduction procedures are identified that may allow for similar small aperture facilities, constructed from commercially available/off-the-shelf components, to improve image and subsequent data quality.
    A demonstration of the hardware and software capabilities is expressed through observation statistics from a 10 month observing campaign, and through the photometric characterisation of near-Earth Asteroids 8014 (1990 MF) and 19764 (2000 NF5).

\end{abstract}

\section{Introduction} \label{sec:introduction}
    
    Despite their limited light gathering power, small aperture ($<\,0.5\,$m) telescopes can contribute widely across all disciplines within astronomy.
    These facilities have the advantage that the rate at which data can be collected, and their availability for short-notice targets of opportunity, is much better than that of larger facilities which are limited by competition for observing time.
    Examples of the many observational projects for which small telescopes make a valuable contribution are: rapid-response gravitational wave follow-up from LIGO/VIRGO alerts \citep[e.g.,][]{2017GCN.21748....1R}, \textit{Gaia} transient follow-up \citep[e.g.,][]{2020A&A...644A..49M,2020A&A...633A..98W}, long-term variable star characterisation \citep[e.g.,][]{2020MNRAS.493..184E}, and exoplanet transit studies \citep[e.g.,][]{2021NewA...8301477S}.
    
    Asteroids provide many opportunities for small telescopes equipped with a good quality CCD camera.
    To date, over a million asteroids have been discovered, around 550 000 of which have confirmed orbits. The Minor Planet Center\footnote{\url{https://www.minorplanetcenter.net/}} \citep[MPC;][]{1980CeMec..22...63M} is responsible for the identification, designation and orbit computation for all of these objects and maintains the master files of observations and orbits. Small aperture facilities can contribute by submitting astrometric observations to the MPC.
	The light collecting area of small telescopes generally precludes photometric observations of distant asteroids such as Trans-Neptunian Objects or Jupiter Trojans, and observations of small asteroids are possible only when they make relatively close approaches to the Earth, as is the case for near-Earth Asteroids (NEAs).
	
	The JPL Small Body Database\footnote{\url{https://ssd.jpl.nasa.gov/sbdb.cgi}} Browser provides further information, including (many, but not all) reported physical properties derived from observations.
	Light curves obtained from relative photometric observations can help to constrain object spin and shape properties \citep[e.g.,][]{2021MPBu...48...30W}.
	Long-term monitoring of rotational properties using small telescopes can contribute to detections of the YORP effect \citep[e.g.,][]{2007Natur.446..420K,2021AJ....161..112L}.
    Occultations of stars by small bodies are particularly accessible to small telescopes.
    These types of observations provide opportunities to directly measure shape information of targets while only requiring the telescope to be of sufficient size to detect the star being occulted instead of the potentially faint occulting object \citep{2020MNRAS.499.4570H}.
    
    The availability of long periods of observing time on small telescopes creates opportunities for targeted studies of individual asteroid phase curves \citep[e.g.,][]{2021P&SS..19505120H,2021Icar..35714158O}.
	Phase curves describe the reduction in the reduced magnitude of an asteroid with increasing phase angle.
	The phase angle is defined as the angle between the Earth and Sun position vectors from the reference frame of the target object.
	The apparent brightness of an asteroid is determined by its intrinsic properties (mean cross section and geometric albedo) as well as the geometry of observation.
	The absolute magnitude of an asteroid, $H$, is a measure of its intrinsic properties.
	$H$ is defined as the apparent mean (over a rotational light curve) Johnson V magnitude of the asteroid if placed at a heliocentric distance $r = 1$ AU, geocentric distance $\Delta = 1$ AU, and a phase angle of $\alpha = 0$ degrees.
	It is related to the mean observed magnitude $V(r, \Delta, \alpha)$ by
	\begin{eqnarray}
	    H &=& V(r, \Delta, \alpha) - 5\log_{10}\left( r\Delta \right) - \phi(\alpha) \nonumber\\
	    &=& V(1,1,\alpha) - \phi(\alpha)
	\end{eqnarray}
	where $V(1,1,\alpha)$ is the reduced magnitude and $\phi(\alpha)$ is the phase function.
	Phase curves can help to constrain the absolute magnitudes, albedos, and likely taxonomic classifications of asteroids \citep[e.g., ][]{1989aste.book...524,2012Icar..219..283O,2016P&SS..123..117P}, and can also be used to extend the parameter space for classifying and studying asteroid families \citep{2021Icar..35414094M}.
	
	In this paper we characterise the capabilities of the $0.43$ m Physics Innovations Robotic Telescope Explorer (PIRATE) as an example of how small aperture telescopes can be used to obtain asteroid phase curves.
    We demonstrate how high quality calibrated photometry may be achieved for similar telescopes through enhanced data collection and processing techniques.
    Refined methods for asteroid image collection and calibration with small aperture observatories are described in Section~\ref{sec:methods}.
    A new moving object data extraction pipeline for the PIRATE facility is outlined in Section~\ref{sec:pipeline}.
    The methods used for characterising physical properties of asteroids using these observations are outlined in Section~\ref{sec:targetCharacterisation}.
    Section~\ref{sec:results} contains a demonstration of capabilities through observations and phase curve extraction of two NEAs.
    The performance of the observatory and pipeline for asteroid observations is assessed in Section~\ref{sec:observatory}.
    
\section{Observation and Image Calibration Optimisation} \label{sec:methods}

    \subsection{PIRATE}
    
        Since the publication of \citet{2011PASP..123.1177H}, who describe the hardware and operation of PIRATE Mk.~I, the facility has undergone a series of upgrades and has been relocated.
    	The PIRATE Mk.~III facility is located $2390\,$m above sea level at Observatorio del Teide, Tenerife (Minor Planet Center Observatory Code: 954).
    	PIRATE is a combined teaching and research telescope \citep{2018RTSRE...1..127K}, allowing students to interact with and control the telescope in real-time while research observations can be conducted autonomously.
        The PIRATE facility is part of The OpenScience Observatories (hereafter referred to as OSO), the astronomy wing of the award-winning OpenSTEM Labs initiative\footnote{\url{http://stem.open.ac.uk/study/openstem-labs/}}.
        
        The facility consists of; a PlaneWave 17-inch ($0.43\,$m) f/6.8 Corrected Dall-Kirkham (CDK) optical tube assembly\footnote{\url{https://planewave.com/product/cdk17-ota/}}, mounted on a 10Micron GM4000 German Equatorial mount\footnote{\url{https://www.10micron.co.uk/shop/mounts/10micron-gm4000-hps-ii-german-equatorial-mount/}}, housed within a $4.5\,$m Baader Planetarium All-Sky dome\footnote{\url{https://www.baader-planetarium.com/en/domes/baader-allsky-domes-(2.3---6.5-meter).html}}.
        The camera is an FLI ProLine PL16803\footnote{\url{https://www.baader-planetarium.com/en/fli-proline-kaf-16803-ccd-camera.html}} with a KAF16803 CCD (4096 x 4096 pixel array, with 9 micron pixels), giving a Field of View (FoV) of approximately 43 x 43 arcminutes and a pixel scale of 0.63 arcseconds per pixel.
        The CCD has a gain of $1.39$ e$^-$/ADU and a read-noise of $14$ e$^-$.
        The camera is fitted with a nine-position filter wheel containing Johnson-Cousins UBVRI filters, Baader narrow-band (Halpha, SII, OIII) filters, and a Luminance filter.
        The facility runs the \textsc{Abot}\textsuperscript{\texttrademark} control software by Sybilla Technologies\footnote{\url{https://sybillatechnologies.com/}} \citep{2014SPIE.9152E..1CS}.
        The control software requests a sequence of observations each night from the OSO scheduler.
        The OSO scheduler is a bespoke software solution that collates all requested programs, and creates an optimal observing schedule for the whole night.        

    \subsection{Observation Strategy} \label{subsec:observationStrategy}

        Observations of asteroids with PIRATE were conducted as sequences of R and V filter observation pairs, to enable the instrumental colour $(V_{inst.} - R_{inst.})$ of the target to be calculated for each night (needed for absolute calibration, see Section~\ref{subsec:lcCalibration}).
        The long exposure times needed due to the small aperture, and the higher typical rates of motion of NEAs, mean that differential tracking is often required.
        For a given night of observations of an asteroid, an average tracking rate, calculated over the start and end of the observation program, in Right Ascension (RA) and Declination (DEC) is used instead of changing rates throughout the observation program.
        The observation program is the sequence of observations of a single object on a given night.
        This reduces telescope overheads and program complexity.
        The telescope begins each exposure at the provided RA and DEC, tracks the target within the exposure, and then resets to the original coordinate between exposures.
        This provides a stable field of trailed stars in images, with a tracked asteroid moving across the frame from one image to the next.
        This is different to the more common method where differential tracking continues between exposures, and the images keep the tracked object in the centre of each image with a moving and trailed star field.
        
        An asteroid observing program generation and optimisation tool was created for PIRATE.
        The tool uses the \textsc{astroquery} Python package \citep{2019AJ....157...98G} to query the \textsc{JPL Horizons} ephemeris system \citep{2015IAUGA..2256293G}, and calculates the mean RA, DEC, and respective mean rates of motion for the times requested.
        Using the combined rate of motion and program length, the tool ensures that the asteroid does not probe the edges of the field of view (i.e. the asteroid should not move more than 40 arcminutes) during the program.
        If this condition is not met, the program length is adjusted accordingly.
        
        True rates of motion for an object over a program are not constant, so using an average rate will provide some accumulated tracking error.
        The tool initially checks if the expected RMS tracking error over the program is less than 1 arcsecond (typical seeing at Teide Observatory), and if not then the program length is reduced until the condition is met.
        Excessive trail lengths of background stars prevents accurate plate solving of the images as well as photometric calibration.
        The tool checks if the combination of exposure time and rate of motion will produce trails longer than 100 pixels.
        If this is the case, then the exposure time is reduced until the condition is met.
        Once all of these program optimisation steps are completed, the tool sends the request containing RA, DEC, RA rate, DEC rate, filters, exposures, and timings to the OSO scheduler.
    
    \subsection{Optimising Standard Data Reduction Processes} \label{subsec:processOptimisation}
    
        In this subsection we present an investigation into the standard data reduction procedures on PIRATE.
        Bias temperature and structure drift, dark non-linearity, and flatfield variability are identified.
        Potential optimisations are outlined to minimise the noise from these sources in reduced images.
        These investigations and optimisations can be applied to other small aperture facilities in order to minimise noise and maximise scientific output.
        
        \subsubsection{CCD Bias Characterisation} \label{subsubsec:biasCharacterisation}
        
\begin{figure}[t]
                \centering
                \includegraphics[width=\textwidth]{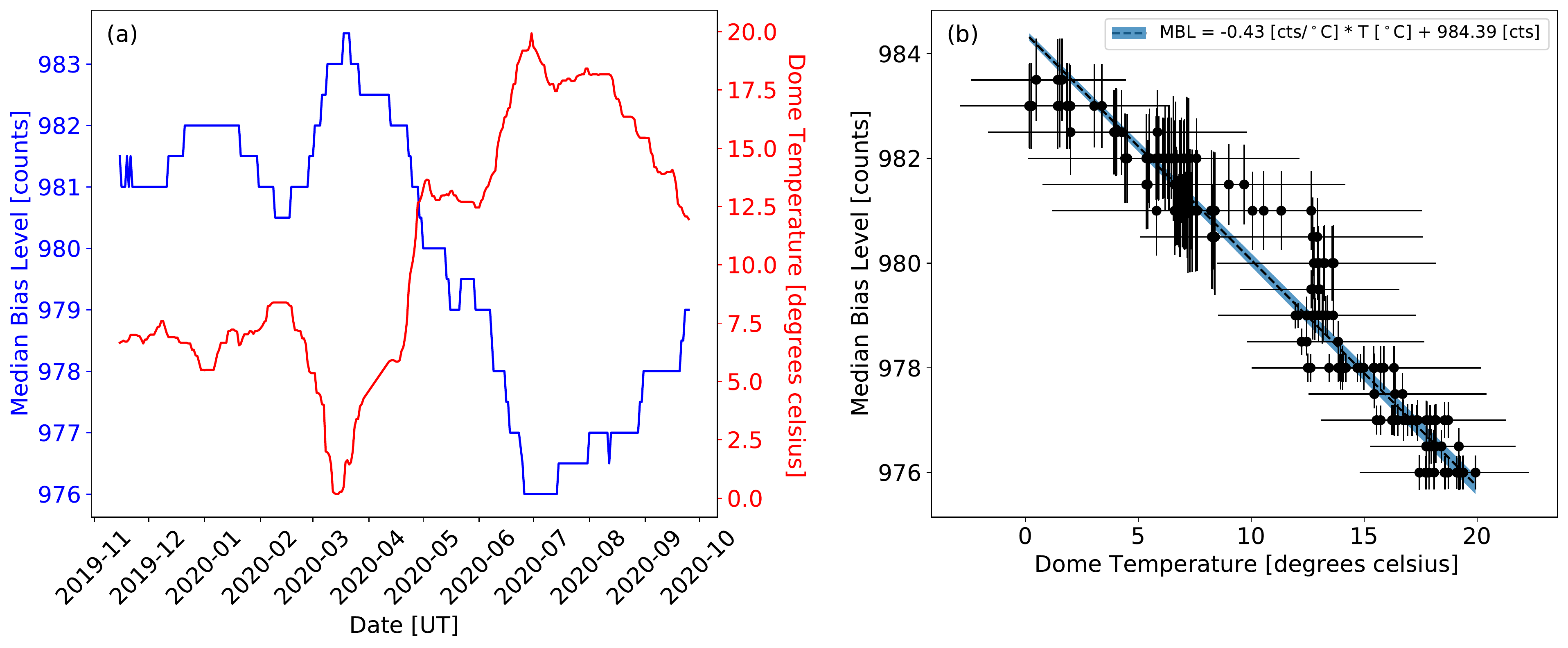}
                \caption{\textbf{(a)} Variation of 28-day rolling average bias level with time (blue), over-plotted with 28-day rolling average of dome temperature at time of bias acquisition (red). \textbf{(b)} 28-day average bias level plotted against corresponding 28-day average dome temperature. Uncertainties are taken as the standard deviation of the data used to create each point. The data are strongly correlated.}
                \label{fig:bias-temp-relations}
            \end{figure}

            In theory the bias voltage of a CCD should not be variable with time or temperature as it is set by the read-out electronics of the camera, and a zero-second exposure does not allow for any accumulation of thermally-excited electrons (i.e. dark current).
            However, upon inspection of just under a year's worth of bias frames collected by PIRATE there is variability observed.
            Variability in the 28-day rolling average of mean bias level is commensurate with the 28-day rolling average of the dome temperature at the time of bias acquisition (Figure~\ref{fig:bias-temp-relations}(a)).
            
            The bias level has a strong negative correlation with dome temperature and a linear relationship can be fit to the data (Figure~\ref{fig:bias-temp-relations}(b)).
            The CCD temperature of each frame is observed to be constant, so is not thought to be the cause of the bias level variations.
            The inverse nature of this temperature dependence rules out the possibility of residual thermal noise before read-out as this would produce an increased bias level with increasing temperature.
            Upon further investigation, no definitive cause of this variation is found. 
            However, it is speculated that this may arise from the camera cooling system and read-out electronics not having separate input power systems.
            The input power to the camera is a single 12V line, which must serve the power needs of both the CCD and other camera functions such as the shutter and Peltier cooler.
            As the temperature rises the cooler requires more power to keep the CCD at its operational temperature of $-30^\circ$C, and therefore the read-out electronics may experience a corresponding dip in power, leading to a corresponding drop in the bias level.
            
\begin{figure}[t]
                \centering
                \includegraphics[width=.6\linewidth]{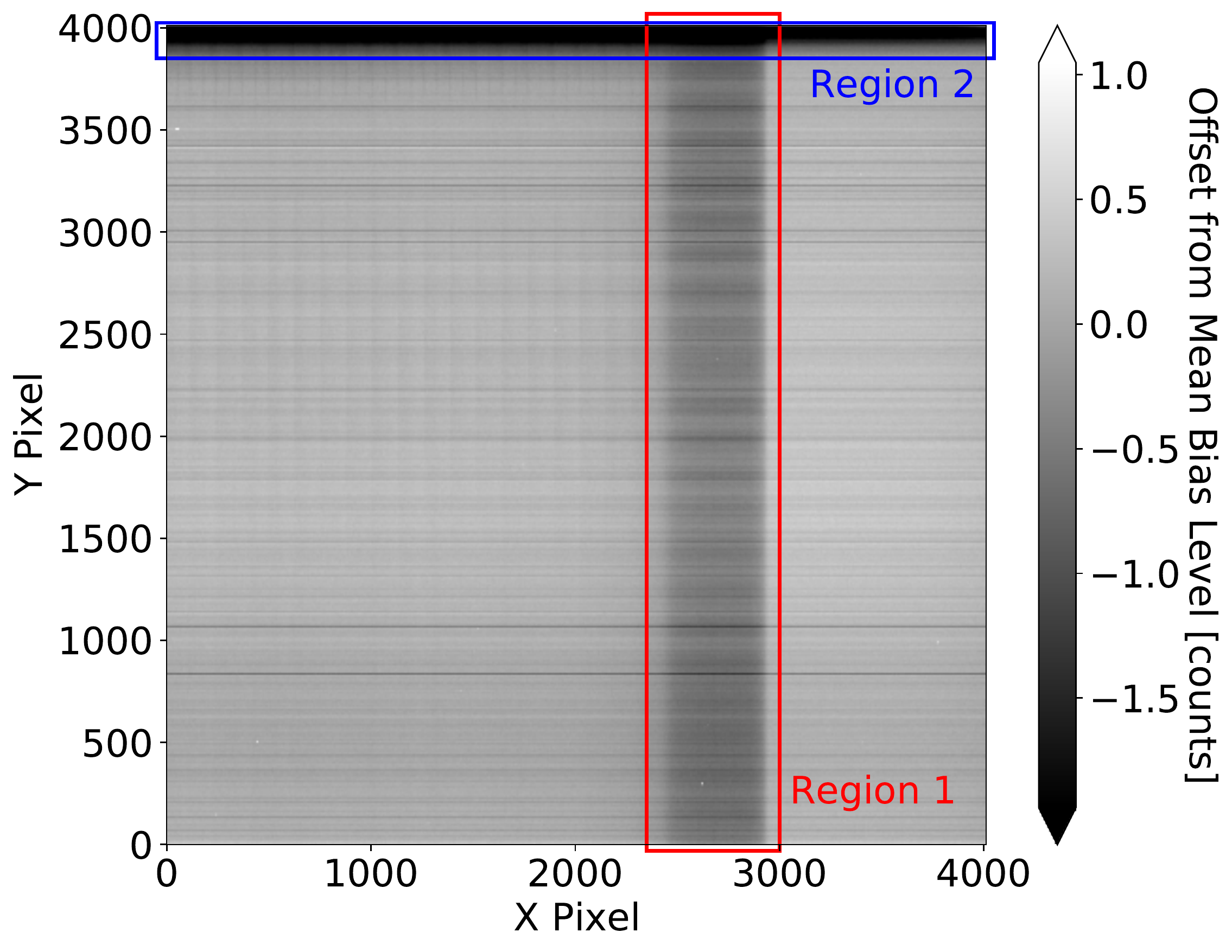}
                \caption{Bias structure map, created from the average of a year of master bias frames with the median bias level subtracted from each.}
                \label{fig:biasStructureMap}
            \end{figure}

            It is expected that the structure of the CCD bias will be stable with time as long as the CCD is not altered or damaged.
            By taking a master bias frame and subtracting the sigma-clipped median bias level from this frame, we obtain a map of the structure of that master bias.
            By repeating this for a year's worth of master bias frames and median combining the structure maps of each, we obtain a bias structure map as shown in Figure~\ref{fig:biasStructureMap}.
            There are two large regions of pixels that seem to have a distinctly lower bias level than their neighbouring pixels.
            To assess the stability of the map, the data used to construct it were split into two periods and the output maps compared.
            The residual map between the two periods indicates that the bias structure is variable in the area corresponding to region 1 in Figure~\ref{fig:biasStructureMap}.
            The bias structure map can therefore not be used over long time periods and is instead generated each month to minimise the effect of variability.
            
            From the structure map we can obtain a synthetic bias image to use when calibrating each flatfield and science frame.
            The structure map is scaled according to the dome temperature obtained from the FITS header of the image to be calibrated.
            This is calculated using the relationship fit to the data in Figure~\ref{fig:bias-temp-relations}(b):
            \begin{eqnarray}
                \textbf{BIAS}_\text{T} = &984.39 - 0.428 \times \text{T} + \textbf{ STRUC\_MAP}.
            \end{eqnarray}
            $\textbf{BIAS}_\text{T}$ is the synthetic bias image (counts) at a dome temperature T ($^\circ$C), $984.335$ counts \& $-0.428$ counts/$^\circ$C are the y-intercept and gradient of the bias-temperature relationship, and $\textbf{STRUC\_MAP}$ is the bias structure map (counts).
            This method provides a significant improvement on the noise level from median combining bias frames from a few nights.
            A reduction in noise to this level using standard methods would require the combination of hundreds/thousands of bias frames, while also still not accounting for the temperature-induced bias drift in each image.
            It is possible for the overscan region of each frame to be used to account for the variation in the bias due to temperature.
            However, this will be noisier than our method which characterises the effect using a large number of frames.
            Using the overscan region to account for this variation would be ideal for cameras that undergo bias variation much larger than the read noise of the CCD.
                
        \subsubsection{Dark Current Characterisation}
            
            Thermal noise on PIRATE is expected to be very low at or below -30$^\circ$C ($< 0.005$ e$^-$/sec/px at -35 degrees according to the CCD specification sheet\footnote{\url{https://www.baader-planetarium.com/en/downloads/dl/file/id/1508/product/4266/overview_fli_proline_ccd_kamera_kaf_16803.pdf}}).
            It is therefore optimal for very long exposure dark frames to be taken to sample the dark current well.
            However, this is not possible to do automatically on PIRATE due to the limited time available during dusk and dawn for calibration frames, and the desire not to occupy dark hours with calibrations.
            The standard procedure on PIRATE, prior to this work, was to obtain a sequence of 60 second darks that are then median combined and scaled to obtain an estimate of the dark current in each image on a given night.
            
\begin{figure}[t]
                \centering
                \includegraphics[width=.6\linewidth]{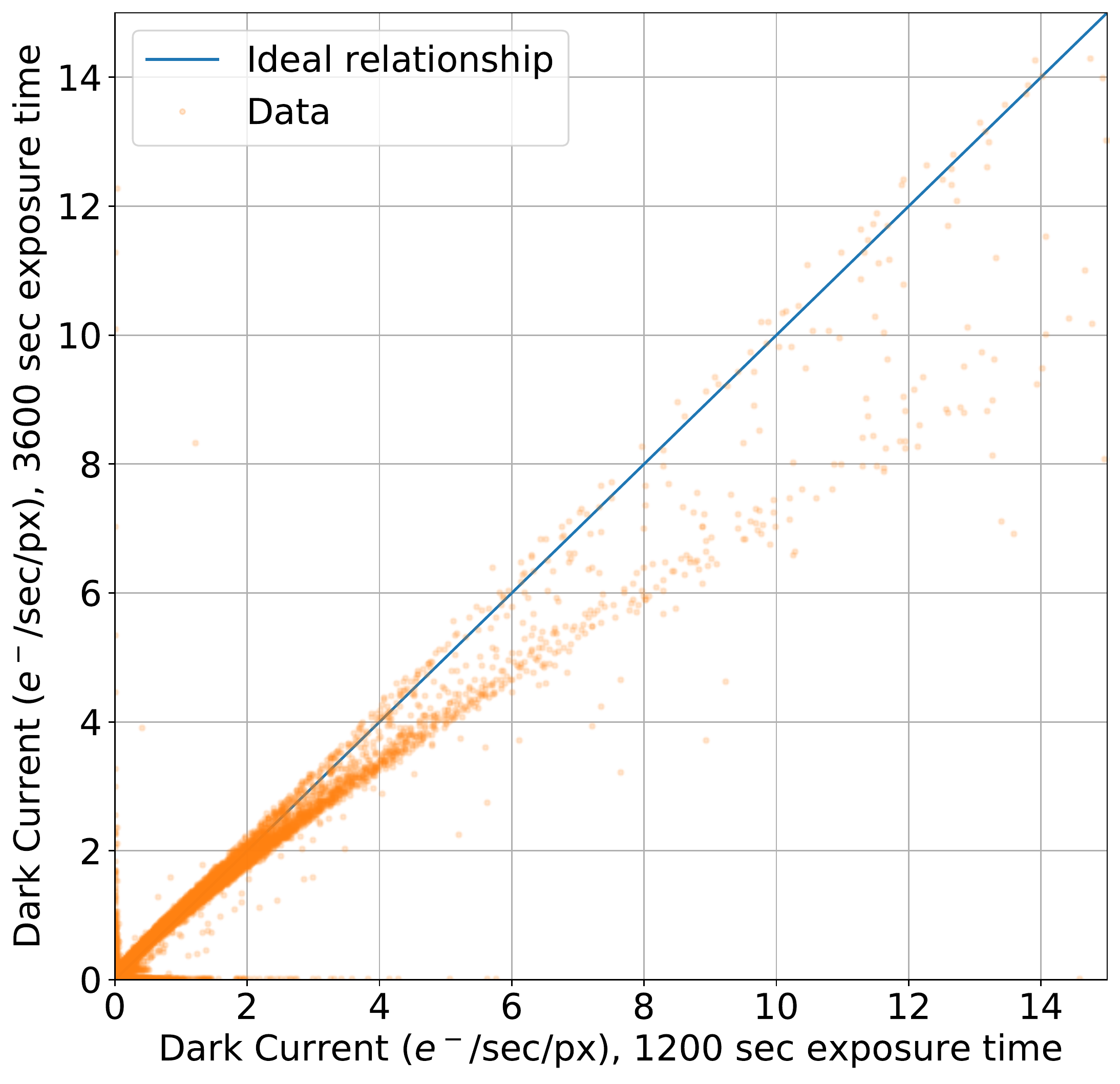}
                \caption{Cross-histogram showing the dark current in each pixel for two different dark frames, with exposures of 1200~s and 3600~s. A clear deviation from the expected 1:1 relationship is observed, limiting the effectiveness of dark correction for these pixels.}
                \label{fig:crossHist}
            \end{figure}

            The median dark current across the CCD is 0.005 $e^-$/sec/px, meaning that in a single 300-second science frame (the largest exposure time typically used to avoid the accumulation of differential tracking errors) the dark current will be significantly less than the typical read noise in the image ($14$ e$^-$).
            To investigate the dark current present in the CCD, two dark frames with very long exposures (1200 \& 3600 seconds) were taken and compared via a cross-histogram (Figure~\ref{fig:crossHist}).
            We expect that pixels will show a consistent level of dark current from frame to frame, and hence a 1:1 relationship is expected in the cross-histogram.
            However, many of the pixels that have a dark current larger than the read noise in typical exposures appear to show a non-linear or variable thermal response with exposure time (this population is referred to as the `hot pixels' hereafter).
            The population of pixels along the edge of either axis can be ignored as these are pixels that have had a cosmic ray hit in one frame, but not the other.
            The scaling issues for high dark current pixels and the overall low dark current across the rest of the CCD means that many pixels are not represented properly by dark frames unless we are able to always match dark exposure lengths to that of all science images.
            This is not efficient for an automated system that handles many types of request over a single night.
            
            The count distributions of the hot pixel population in science images were calculated when calibrating using three different cases: long exposure dark frames, 60 second darks, and without dark correction.
            Under the assumption that calibrating using the long exposure dark frames will be most accurate, we compare them with the count distributions using the 60 second darks and no dark correction cases.
            We find that forgoing dark correction provides a comparable representation of the hot pixels in the image to using the long exposure darks.
            This is not the case when using the 60 second darks, where frequent over- and under-estimations of the dark current in these pixels are observed.
            This indicates that avoiding dark correction on the PIRATE CCD provides a better approximation (for both low and high dark current pixels) to the case with long exposure darks than is possible using the 60 second darks due to the poor sampling capability of these frames.
            Therefore, the standard procedure going forward is to avoid dark correction for PIRATE data that has been taken with the CCD cooled to -30$^\circ$C.
            
            However, the variable/non-linear response of the hot pixels is still an issue in flatfield and science frames.
            To minimise the effects of these pixels on our data, a map of pixel coordinates was generated by identifying those that are consistently in the `dark non-linearity region' of $>$ 0.5 $e^-$/sec/px over time.
            This map can then be used as an interpolation map for flatfield and science frames, where each pixel coordinate in the map is replaced by an average of its neighbours.
            
        \subsubsection{Flatfield Characterisation}
            
            The construction of a high quality flatfield is of particular importance with moving objects that occupy many different pixel locations throughout the night.
            It is almost impossible to get a truly flat input light source, with this becoming increasingly difficult as the telescope field of view increases.
            PIRATE flatfields are taken at twilight by pointing the telescope at a position close to the null spot of the twilight sky \citep{1996PASP..108..944C}, although this is an approximation as there is no point on the twilight sky which is uniformly flat.
            Another issue is that the pixel-to-pixel variations of the CCD are wavelength dependent, and so it is optimal for the wavelength distribution of the flatfield source to be the same as in the science images \citep[][p.~69]{2006hca..book.....H}.
            The spectrum of the twilight sky is different from that of the night sky and the observed sources in the image, and so the pixel-to-pixel variations determined from twilight flats are at best an approximation to those present at night.
            The spectral difference is not something that can be counteracted by changes to operational processes, and is therefore not characterised in this work.
            
            The effects caused by slight gradients in the twilight sky, however, may be observable over time if the telescope consistently points in the same place for flatfields each night.
            Due to the changing azimuth of the Sun when it sets throughout the year, the gradient across the twilight sky at a given altitude and azimuth will change.
            This could cause slight long-term shifts in the gradient across the chip over time.
            Differing weather conditions and atmospheric dust content (a common issue in the Canary Islands) from night to night could also introduce variability into the flatfield frames, through the alteration of sky gradient or spectrum.
            
            In order to produce a high quality flatfield, frames are combined from multiple nights to reduce the noise level.
            However, sources of variability must be adequately characterised to understand their timescales and magnitudes so that we do not introduce too much noise from these effects into the master flatfield.
            To assess the variability of the structure of these frames, we define the `centre-to-corner' ratios.
            These are calculated as the ratio between the average of a central region ($\sim 160000$ pixels) in the flatfield and the average of an area of equivalent size in each of the four corners.
            A rapidly changing `centre-to-corner ratio' indicates a high level of variability in the structure of the flatfield.
            
\begin{figure}[t]
                \centering
                \includegraphics[width=\textwidth]{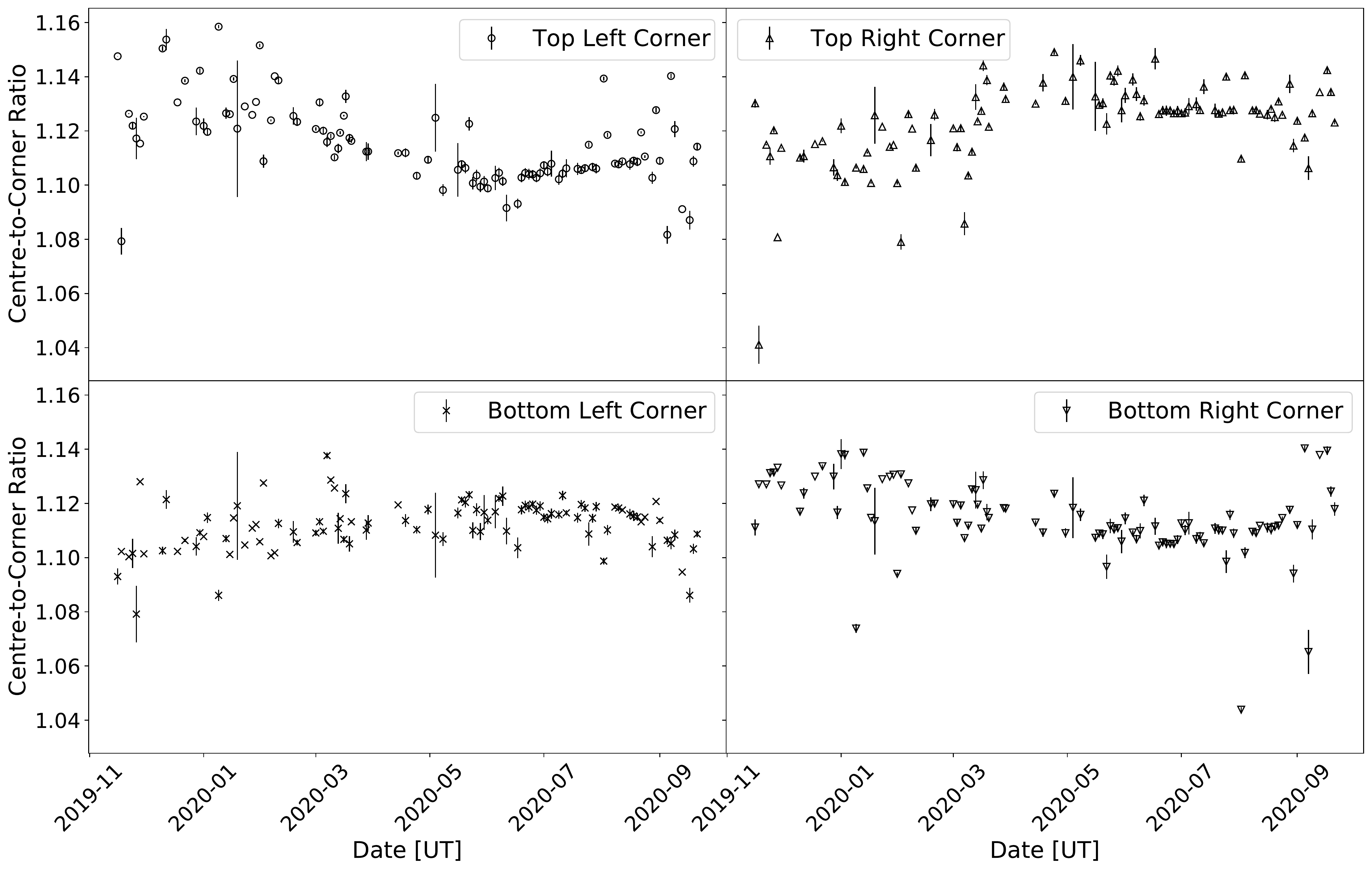}
                \caption{Variation of flatfield centre-to-corner ratios with time for the R filter flatfields. Uncertainties represent the standard deviations of the intra-night ratios used to calculate the mean, in order to give an idea of the scatter within a night.}
                \label{fig:nightToNightVariations}
            \end{figure}

            The first step was to characterise the intra-night and night-to-night variations in the flatfields (i.e. the short-term variability).
            To do so, the corner ratios from each individual flatfield on a given night were calculated, and the mean and standard deviation recorded.
            The variations with time of these ratios in the R filter are shown in Figure~\ref{fig:nightToNightVariations}, with the uncertainties representing the standard deviation of the intra-night ratios.
            The average night-to-night percentage variation was calculated across all four corners, and across the four filters important for research data collection (B, V, R, I), as $\sigma_{\text{N-N}} \approx 0.51\%$.
            The average of the intra-night standard deviations (flat-to-flat variations) is $\sigma_{\text{F-F}} \approx 0.16\%$, showing that the short-term variability in the flatfields is dominated by the night-to-night variations.
            
            If the short term variability is caused by atmospheric effects, then the flatfield could be changing throughout the night.
            This means that any flatfield created will not be accurate, although combining over many nights allows us to create a flatfield based on average conditions at the site that will represent the flatfield, on average, during the night as well.
            If the short term variability is due to changes to the intrinsic flatfield (e.g. slight shifts in optical alignment) then the best solution would be to only use flatfields from the same night.
            The source of these night-to-night variations is most likely due to changing atmospheric conditions as discussed previously, as a single night is not long enough for significant changes to solar-induced sky gradient or instrumental changes to occur.
            
            To analyse the long-term variations in the flatfields, the data were regenerated using a 28-day rolling average of the flatfields to smooth out the short-term variability.
            Temperature, solar azimuth, and solar angular separation from telescope pointing at the time of flatfield acquisition have been investigated as causes of these long-term variations.
            However, correlations between any of these parameters and the flatfield variations are absent.
            This indicates a possible intrinsic flatfield change, perhaps due to slight movement of optical components in the telescope system over long timescales.
            
\begin{figure}[t]
                \centering
                \includegraphics[width=.6\linewidth]{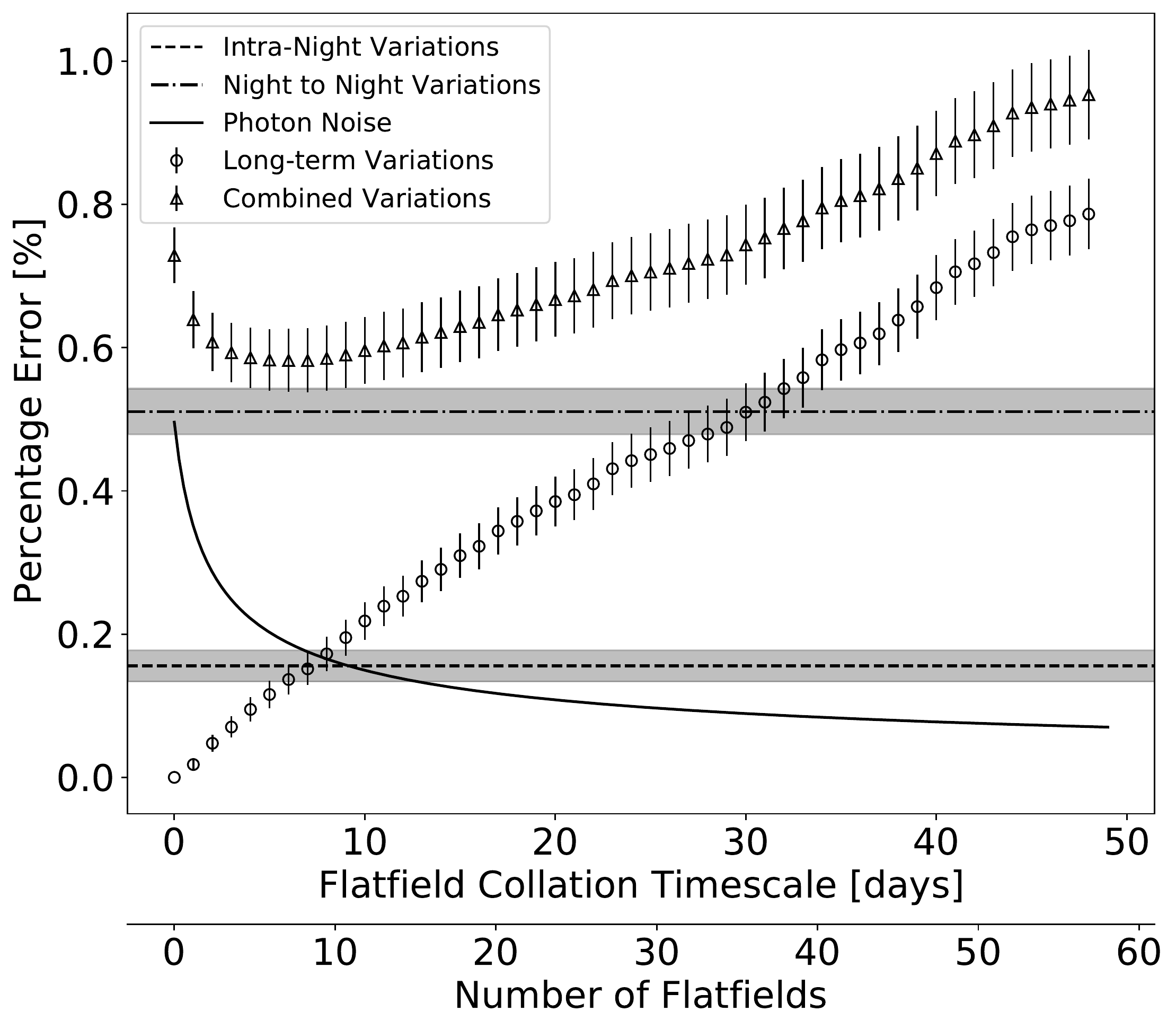}
                \caption{Estimated percentage error in flatfields introduced by various flatfield noise sources. The dashed line represents the error introduced by intra-night variations, the dotted \& dashed line represents the percentage error introduced by night-to-night variations in the flatfields. The solid line represents the reduction in photon noise in the pixel-to-pixel variations with additional flatfields. Open circles represent the increasing error with increased flatfield collation time introduced by long-term flatfield variations. Open triangles are all noise sources combined in quadrature to obtain estimated total flatfield error. The optimal flatfield collation time is where the combined noise is minimised (approximately 8 days).}
                \label{fig:flatfieldCollationTimescale}
            \end{figure}

            This long-term variability constrains the timescale over which flatfield frames can be collated.
            For a range of collation timescales between 0 and 50 days, the average percentage error from long-term variations was calculated from the data (open circles, Figure~\ref{fig:flatfieldCollationTimescale}).
            The pixel-to-pixel variation error (solid line, Figure~\ref{fig:flatfieldCollationTimescale}) is calculated as the reduction in photon noise expected when using flatfields with a measured average count level of $\sim$34000 counts, with an average of 1.18 usable flats collected per day on average in each filter.
            This average includes nights where no flatfield frames were acquired due to poor weather to reflect the true rate of acquisition.
            The dash-dotted line in Figure~\ref{fig:flatfieldCollationTimescale} represents the night-to-night variations, $\sigma_{\text{N-N}}$, and the dashed line represents the flat-to-flat variations, $\sigma_{\text{F-F}}$.
            The combination of these noise sources (open triangles, Figure~\ref{fig:flatfieldCollationTimescale}), displays a minimum at approximately 8 days of collation time.
            Below this level of collation, pixel-to-pixel variations are not constrained enough, and above this level of collation the long-term variations begin to dominate and introduce additional error in the flatfield.
            
\begin{figure}[t]
                \centering
                \includegraphics[width=.6\linewidth]{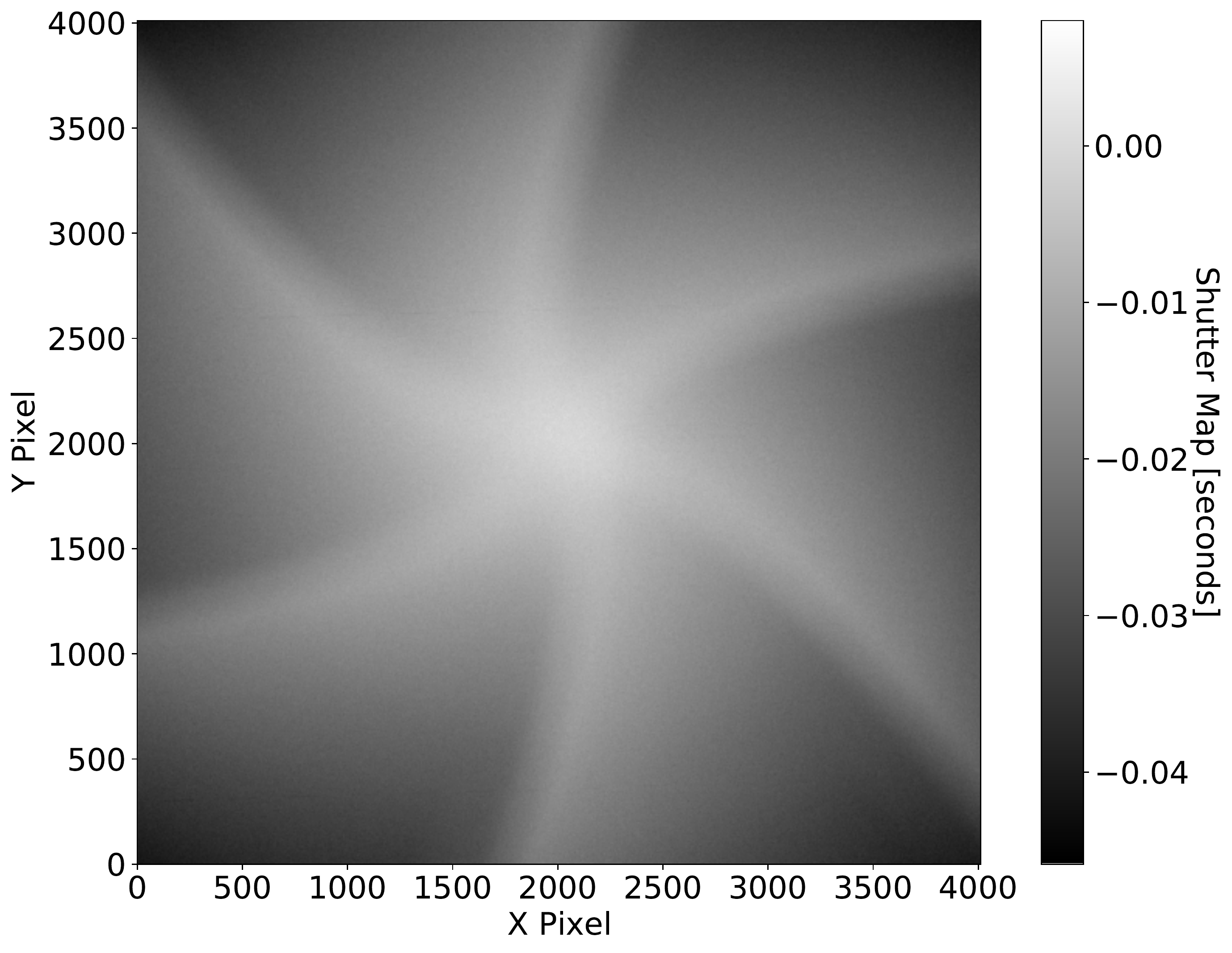}
                \caption{Shutter map created from flatfield frames according to the procedure in \citet{1993A&A...278..654S}. This shutter map can then be scaled using the target frame exposure and used to remove any shutter pattern present.}
                \label{fig:pirateShutter}
            \end{figure}

            There also exists a significant shutter pattern in frames with exposures less than approximately 4 seconds.
            To minimise these effects on cameras with similar iris shutters, frames below these exposure times can be removed from the flatfield stack.
            However, this has the effect of reducing the number of flatfields available for master flatfield creation and may not be ideal for systems where flatfield variability is too high to reliably use flatfields from many previous nights.
            To reduce the shutter effect in these images on PIRATE, a map of the shutter pattern was created following the procedure outlined in \citet{1993A&A...278..654S}.
            The data to create this shutter map were originally collected and processed by \citet{morrellPhD}, and subsequently re-processed as part of this work.
            This shutter pattern is shown in Figure~\ref{fig:pirateShutter}.
            The shutter map can imprint an error on the flatfield greater than 1\% at the dark edges for exposures less than 2.5 seconds.
            The correction to the flatfield using this shutter map reduces the effect of the shutter to less than 0.4 percent for a 2.5 second flatfield, i.e. it reduces the error to a level below that expected from the flatfield variability discussed previously.
            The shutter correction is applied to images as follows, from a re-arrangement of Equation~2 in \citet{1993A&A...278..654S}:
            \begin{equation}
                \mathrm{\textbf{CI}} = \frac{\mathrm{\textbf{UI}}}{1 + \frac{\boldsymbol\beta}{\boldsymbol\alpha}\frac{1}{t}},
            \end{equation}
            where \textbf{CI} is the corrected image, \textbf{UI} is the uncorrected image, $\frac{\boldsymbol\beta}{\boldsymbol\alpha}$ is the shutter map (Figure~\ref{fig:pirateShutter}), and $t$ is the exposure time of the image.
            
\section{Asteroid Data Extraction Pipeline} \label{sec:pipeline}

    \subsection{Light Curve Extraction} \label{subsec:lcExtraction}
    
        In order to sustain the required data collection cadence for asteroid phase curve studies, automated tools for data extraction are required.
        As part of the standard data reduction pipeline for PIRATE, all frames undergo plate solving using a local installation of \textsc{Astrometry.net} \citep{2010AJ....139.1782L}.
        The images subsequently undergo source extraction using \textsc{Source Extractor} \citep{1996A&AS..117..393B}, which outputs a catalogue for each image with instrumental photometry and various measured properties of all sources.
        The images in this work contain both star trails and tracked asteroids, so the \textit{MAG\_AUTO} measurement is used as it uses an elliptical aperture that automatically scales to contain approximately $90\%$ of the source flux.
        This ensures consistent measurements between trails and point sources as the shape of the ellipse can be easily tailored to either, while still containing approximately the same flux percentage in the aperture.

        A new automated data extraction pipeline for moving object images from PIRATE has been developed that uses the plate solved images and associated \textsc{Source Extractor} catalogues.
        The benefit of having plate solved images prior to running \textsc{Source Extractor} is that each source can be identified by its equatorial coordinates instead of using pixel coordinates.
        This allows for automated target identification by querying \textsc{JPL Horizons} for the target coordinates at the time of each exposure.
        The output \textsc{Source Extractor} catalogues can then be queried to identify if there are any objects within a defined radius of that coordinate (allowing for plate solving and orbit solution uncertainties).
        
        Using these catalogues, stars that appear in every frame are identified to use as an ensemble of reference stars.
        We identify a reference frame in each filter, with the reference frames chosen as the R and V frame pair that has the lowest airmass.
        For each reference star, the shift of its instrumental magnitude from its value in the reference frame is measured.
        Measuring the weighted average of this shift in each frame for the entire ensemble of reference stars allows us to measure the extrinsic brightness variations in the images due to changing airmass or weather, for example.
        We then correct the target instrumental magnitudes in each frame for these extrinsic variations, leaving just the intrinsic brightness variations of the target (i.e. the relative light curve).
        
        This process is done independently for each filter (V and R) to get a relative light curve in each filter.
        At the observation time of each point on the V light curve, we use linear interpolation of the R filter light curve to estimate the R magnitude at that time, and measure the difference between the V and R magnitudes for all points on the V light curve.
        From these values we determine the average instrumental colour, $\overline{(V_{inst.} - R_{inst.})}$, and corresponding uncertainty (taken as the standard deviation).
        This colour is needed for absolute calibration of the light curve, which will be discussed further in Section~\ref{subsec:lcCalibration}.
        As we will be transforming instrumental R-band magnitudes to true magnitudes in the calibration process, we do not require a separate instrumental V-band light curve.
        We can therefore adjust the V filter light curve to overlap with the R filter light curve using this average instrumental colour, $\mathbf{R_{V}} = \mathbf{V} - \overline{(V_{inst.} - R_{inst.})}$.
        The final relative lightcurve is the combination of the relative light curve in the R filter, combined with the shifted light curve in the V filter,
        \begin{equation}
            \mathbf{R_{frame}} = \mathbf{R} \cup \mathbf{R_{V}}
        \end{equation}
        This process leaves us with a single relative light curve in the instrumental R filter (hereafter referred to as $\mathbf{R_{frame}}$) and a measurement of the average instrumental colour.
        
    \subsection{Light Curve Calibration} \label{subsec:lcCalibration}
    
        To transform the relative light curves into true (apparent) magnitudes, the pipeline adopts the process described in \citet{2017MNRAS.471.2974K}, using the Pan-STARRS photometric catalogue \citep{2020ApJS..251....7F} to calibrate the measurements.
        This catalogue is chosen due to the density of available sources at magnitudes similar to those observable with PIRATE, the easily queried API, and the availability of colour transformation equations back into the Johnson photometric system.
        Asteroid phase curves and associated models are defined in the Johnson V-band, and so once these observations are calibrated to the Pan-STARRS photometric system, they must be converted into the Johnson V-band.
        It would be preferable to use a catalogue in the V-band with many high quality sources and an easily queried API in order to reduce the number of conversions required.
        However, no such catalogue is found that fits the criteria.
        
\begin{figure}[t]
            \centering
            \subfloat[]{ \includegraphics[width=0.49\textwidth]{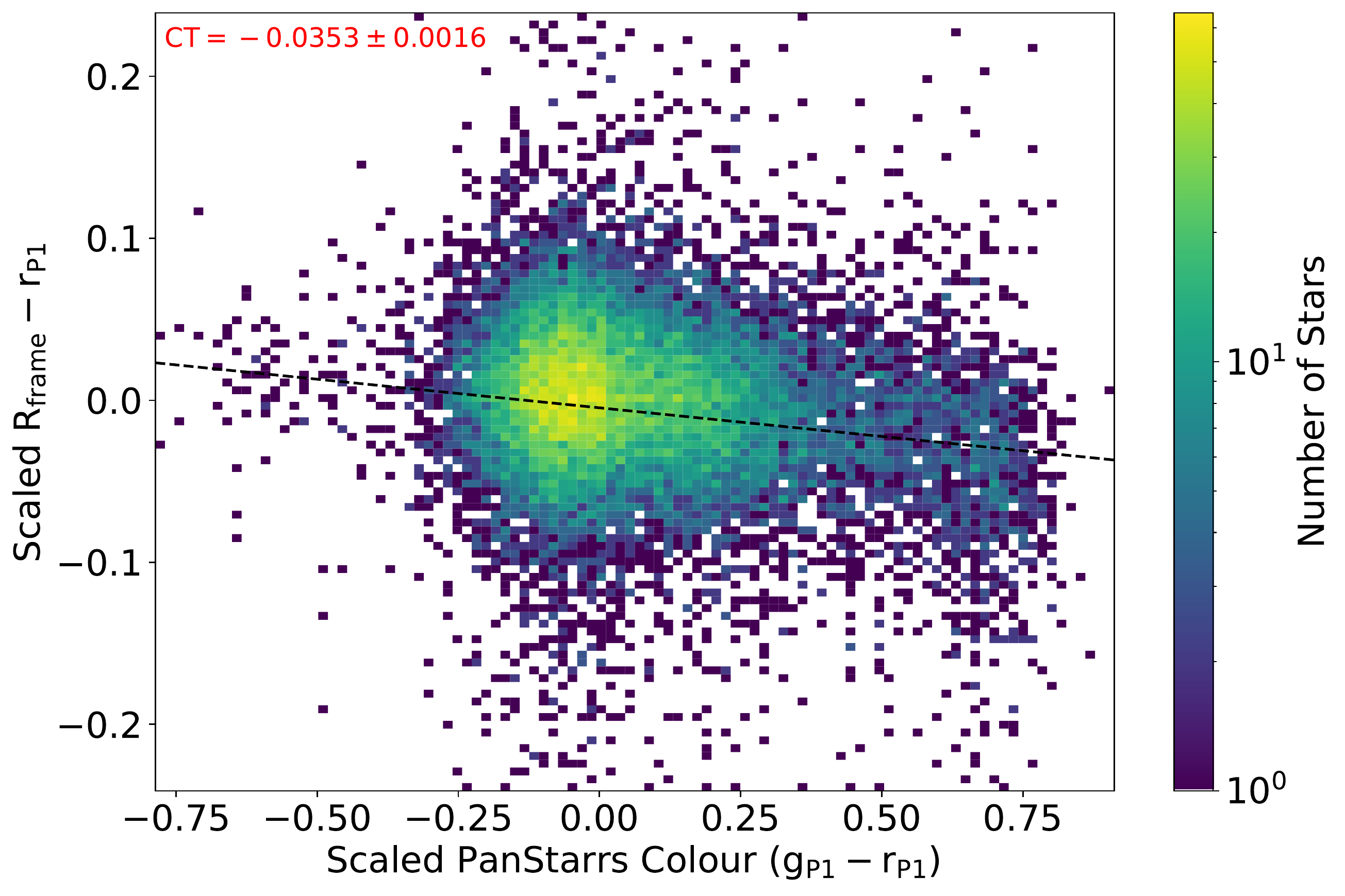}}
            \subfloat[]{\includegraphics[width=0.49\textwidth]{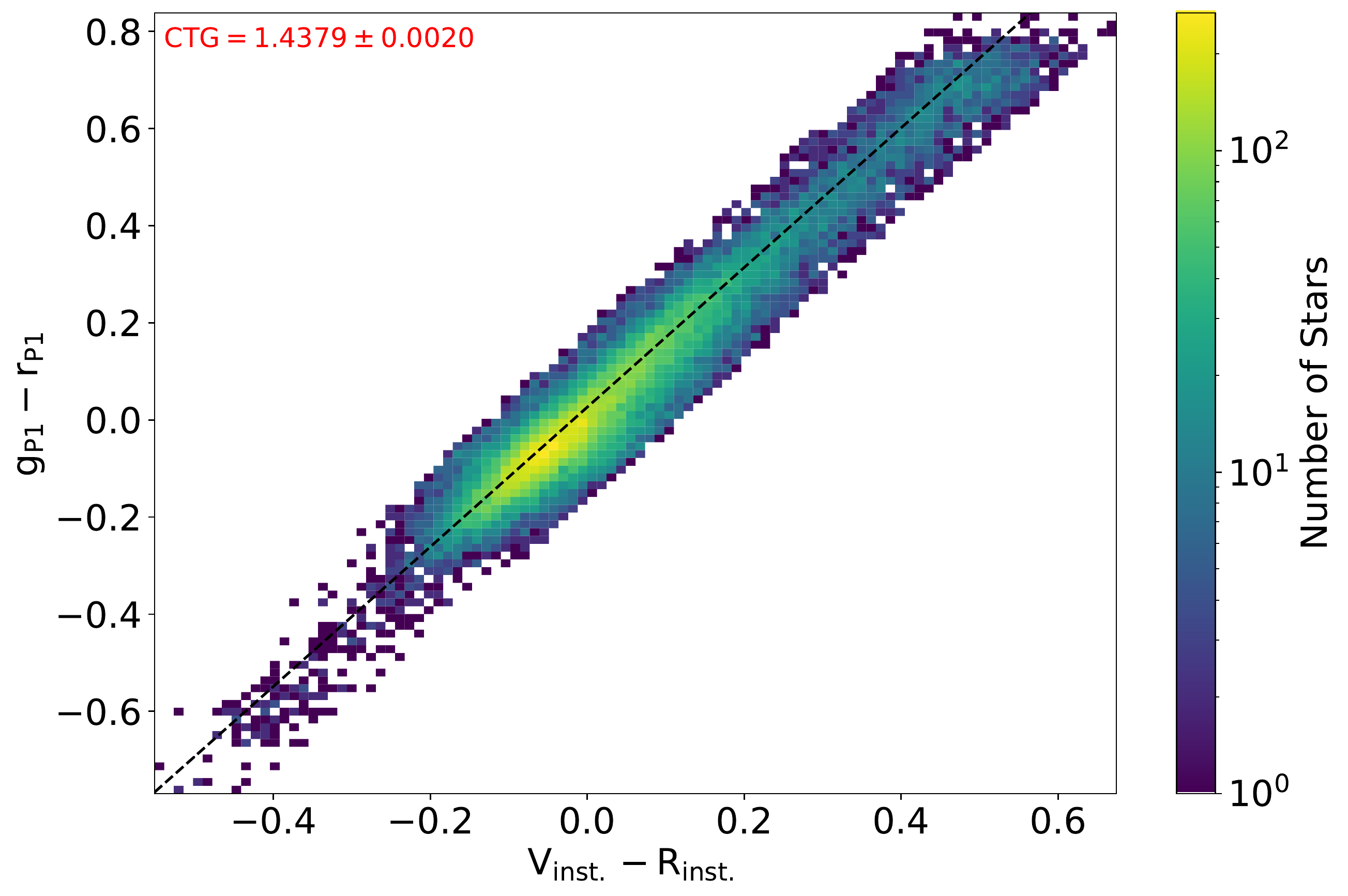}}
            \caption{\textbf{(a)} Difference between instrumental R and Pan-STARRS $r_{P1}$ magnitudes against Pan-STARSS colour. Data collated over many nights and scaled to a mean of zero in each axis for each data set. Colour Term (CT) is measured as the gradient of the best fit linear function, $CT = -0.0353 \pm 0.0016$. \textbf{(b)} Pan-STARRS $(g_{P1} - r_{P1})$ colour against instrumental $(\vr)$ colour. Data collated over many nights and scaled to a mean of zero in each axis for each data set. Colour Transformation Gradient (CTG) is measured as the gradient of the best fit linear function, $CTG = 1.4379 \pm 0.0020$.}
            \label{fig:colourTermandTransform}
        \end{figure}

        The transformation from instrumental R-band magnitudes to the Pan-STARRS $r_{P1}$ system is achieved through a linear colour correction (Equation~\ref{eq:kokotanekovaCalibration}).
        \begin{equation} \label{eq:kokotanekovaCalibration}
            \mathbf{r_{P1}} = \mathbf{R_{frame}} - CT\left(g_{P1}-r_{P1}\right) - ZP
        \end{equation}
        $\mathbf{r_{P1}}$ is the Pan-STARRS r-band light curve of the target, and $\mathbf{R_{frame}}$ is the relative light curve of the target as outlined in Section~\ref{subsec:lcExtraction}.
        The colour term, $CT$, describes the colour correction of the instrumental R filter to the Pan-STARRS $r_{P1}$ filter.
        By plotting the difference between instrumental R-band and Pan-STARRS $r_{P1}$ magnitudes of stars against their corresponding Pan-STARRS $(g_{P1} - r_{P1})$ colours from many frames over a long time-base, and normalising the data in each axis for each frame to a mean of zero, we can measure the colour term as the gradient of the best fit linear relationship.
        This is measured as $CT = -0.0353 \pm 0.0016$ for PIRATE (Figure~\ref{fig:colourTermandTransform}(a)).
        This value is expected to be stable over long time periods, as it is solely dependent on the instrumental set-up.
        $ZP$ describes the nightly zero point correction between the two systems, and depends on atmospheric conditions.
        $\left(g_{P1}-r_{P1}\right)$ is the Pan-STARRS colour of the target, which needs to be measured for each night.
        
        If we fit a linear relationship between instrumental colour $\left(V_{inst.}-R_{inst.}\right)$ and Pan-STARRS colour $(g_{P1}-r_{P1})$ of stars in the images (Equation~\ref{eq:colourTransform}), the colour transformation intercept ($CTI$) will vary from night to night, but the gradient is expected to be stable over time and is measured using an ensemble of frames in a similar way to the colour term.
        The gradient is measured as $CTG = 1.4379 \pm 0.0020$ (Figure~\ref{fig:colourTermandTransform}(b)).
        \begin{equation}
            (g_{P1} - r_{P1}) = CTG \left( V_{inst.} - R_{inst.} \right) + CTI \label{eq:colourTransform}
        \end{equation}
        Using the reference frames for each filter, we can measure the instrumental colour of each star in the field, and plot these against the corresponding Pan-STARRS colour and measure the CTI for the night.
        This means the calibration equation can be re-written as follows:
        \begin{equation}
            \mathbf{r_{P1}} = \mathbf{R_{frame}} - CT\left[CTG\overline{\left(V_{inst.} - R_{inst.}\right)} + CTI\right] - ZP. \label{eq:rP1}
        \end{equation}
        
        As previously mentioned, we need phase curves in Johnson V, and this can be obtained using a transformation equation between Pan-STARRS $r_{P1}$ and Johnson V from \citet{2012ApJ...750...99T},
        \begin{equation} \label{eq:VbandFromRp1}
            \mathbf{V} = \mathbf{r_{P1}} - 0.082 + 0.462\left( B - V \right) - 0.041\left( B - V \right)^2.
        \end{equation}
        We therefore also need to obtain the $(B - V)$ colour of the asteroid. Using equations from \citet{2012ApJ...750...99T} once again, it can be obtained from the Pan-STARRS colour (and subsequently the instrumental colour) as follows:
        \begin{subequations}
        \begin{eqnarray}
            \left(g_{P1}-B\right) =& -0.104 - 0.523\left(B-V\right), \label{eq:gP1toB}\\
            \left(r_{P1}-V\right) =& 0.077 - 0.415\left(B-V\right). \label{eq:rP1toV}
        \end{eqnarray}
        \end{subequations}
        Combining Equations~\ref{eq:gP1toB}, \ref{eq:rP1toV}, and \ref{eq:colourTransform}, the $(\bv)$ colour is
        \begin{subequations}
        \begin{eqnarray}
            \left(\bv\right)&=&\frac{\left(g_{P1}-r_{P1}\right)+0.181}{0.892}, \\[1em]
            \left(\bv\right)&=&\frac{CTG\overline{\left(V_{inst.} - R_{inst.}\right)} + CTI + 0.181}{0.892}. \label{eq:BV}
        \end{eqnarray}
        \end{subequations}
        Substituting our formula for $r_{P1}$ (Equation~\ref{eq:rP1}) and for $\left(\bv\right)$ (Equation~\ref{eq:BV}) into Equation~\ref{eq:VbandFromRp1}, we obtain the V magnitude as follows:
        \begin{eqnarray}
            \mathbf{V} =& \mathbf{R_{frame}} - CT\left[CTG\overline{\left(V_{inst.} - R_{inst.}\right)} + CTI\right] - ZP \nonumber\\
            & - 0.082 + 0.462\left[\frac{CTG\overline{\left(V_{inst.} - R_{inst.}\right)} + CTI + 0.181}{0.892}\right] \nonumber\\
            & - 0.041\left[\frac{CTG\overline{\left(V_{inst.} - R_{inst.}\right)} + CTI + 0.181}{0.892}\right]^2.
        \end{eqnarray}
        
        The relative light curve ($\mathbf{R_{frame}}$), nightly zero point ($ZP$), the colour transformation intercept (CTI), and the average instrumental colour $\overline{\left(V_{inst.} - R_{inst.}\right)}$ are measured for each night/light curve.
        If there is an error in the determination of the zero point for the reference frame (i.e. from a poor fit), then the calibration will be slightly offset.
        To automatically minimise such offsets; the zero points for all R-filter images are measured, de-trended for extrinsic brightness variations using the weighted average shift of the reference stars, and the average and standard deviation of these data taken as the nightly zero point and corresponding uncertainty.
        The colour transformation gradient (CTG) and the colour term (CT) are well defined from an ensemble of observations as described previously.
        The V-band light curve contains uncertainties from two individual components: the photometric uncertainty on each data-point $\sigma_{\mathbf{R_{frame}}}$, and the scale uncertainty arising from the colour and transformation components afterwards, $\sigma_{scale}$ (constant across the whole light curve).
        
\section{Photometric Characterisation Methods} \label{sec:targetCharacterisation}

    To construct high quality phase curves from targeted observations, an analysis of the rotational properties of the object must first be undertaken so that these rotational brightness variations are not imprinted on the phase curve.
    Section~\ref{subsec:spin} details the process used to analyse the rotational brightness variations of target asteroids and the process to account for these variations in the phase curves.
    Section~\ref{subsec:phase} then details the processes and models for extracting phase curves, and methods for characterising targets based on these observations.

    \subsection{Spin-State Analysis and Rotational Averaging} \label{subsec:spin}
    
        To achieve the best quality phase curve the mean brightness of the asteroid is required, and therefore the data must be corrected for incomplete light curve coverage.
        The first step in this process is to get an initial, coarse, estimate of the rotation period of the target.
        This can come from many different sources: a visual inspection of the light curves, prior published data, lomb-scargle periodograms, or a coarse period scan using \textsc{convexinv}\footnote{Available from \url{https://astro.troja.mff.cuni.cz/projects/damit/pages/software_download}.} \citep{2001Icar..153...37K,2010A&A...513A..46D}.
        The last option is usually preferred, and estimates using other data or visual inspection are used in cases where \textsc{convexinv} struggles to find an approximate solution (usually when the time-base of observations is very short).
        After this initial period is found, a rotation model is fit to the reduced magnitude data (i.e. accounting for observer and heliocentric distances) corrected to zero phase angle using an assumed initial linear phase function.
        This rotation model takes the form of a fourth order Fourier series.
        This provides a reasonable fit to light curves without introducing spurious variations due to over-fitting.
        We note that in cases where an opposition surge is present in the data, the period scan using the linear phase function is only conducted on data outside of opposition. Where there is no observed opposition surge, all data are included in the period search.
        
        This is a very rough estimate largely due to the assumed linear phase parameter not properly representing the phase angle variations in brightness.
        To account for this we adopt a linear phase function combined with the fourth order Fourier series.
        Using the estimated Fourier components $A_n$ and $B_n$, we then search a period ($P$) vs. linear phase parameter ($\beta$) grid to find the optimal ($\chi^2_{min.}$) parameter pair that fits the reduced magnitude data.
        The uncertainty is calculated as the range over which the period has a $\chi^2$ value within 10 percent of the minimum $\chi^2$ value.
        
        This process outputs a best estimate for both the period and linear phase parameter.
        These parameters can then be input back into the combined linear-phase Fourier series model to re-fit the Fourier components to better constrain the rotation.
        This process can then be iterated to make gradual improvements to the period and Fourier components.
        This process is more convoluted than some existing methods, but using the calibrated data in this manner is preferable for high amplitude, long-period objects where a full rotation may not be covered in each observation.
        This is due to the requirement for fitting arbitrary magnitude shifts between nights when using uncalibrated, relative, light curves.
        This can introduce over- or under-estimations of the light curve amplitude for these types of slow-rotating elongated objects.
        
        Phase curves only require the average reduced magnitude for each night, so this can be obtained by subtracting the reduced magnitude light curves by the time-dependent part of the Fourier series,
        \begin{eqnarray}
            \mathbf{\overline{V(1,1,\alpha)}} = \mathbf{V(1,1,\alpha, t)} &- \sum_{i=1}^{i=4} \left[ A_n\sin{\frac{2n\pi}{P}\left( t - t_0 \right)} + B_n\cos{\frac{2n\pi}{P}\left( t - t_0 \right)} \right].
        \end{eqnarray}
        The mean of the data is then taken as the average reduced magnitude for the night.
        The uncertainty of the average is then calculated as the standard deviation of the data (i.e. the scatter around the model fit) added in quadrature with the scale uncertainty of the light curve.
        For objects where the rotational variations are too low to be detected by PIRATE, a simple average of the night's data is taken without any rotational modelling and the uncertainty is calculated in the same way as before.
    
    \subsection{Phase Curve Fitting and Classification} \label{subsec:phase}
    
        After the average reduced magnitudes are extracted they are plotted as a function of phase angle and selected models are fit to the data.
        Four models are chosen: a linear phase relation ($H$, $\beta$), the ($H$, $G$) model \citep{1989aste.book...524}, the ($H$, $G_{12}$) model \citep{2010Icar..209..542M}, and the ($H$, $G1$, $G2$) model \citep{2010Icar..209..542M}.
        The model parameters and corresponding uncertainties are estimated from the data using Markov Chain Monte Carlo techniques implemented using the \textsc{emcee} Python package \citep{2013PASP..125..306F}.
        Prior distributions are uninformative, except for setting limits on the parameter ranges to avoid converging to non-physical solutions.
        The model parameter estimates and corresponding uncertainties are calculated as the means and standard deviations of the marginalised posterior sample distributions.
        
        The scatter of the data around the models is mostly driven by: the inability of the models to fully represent the properties of the asteroid (these are only semi-physical models), slight calibration errors, and rotational modelling uncertainties.
        To help reduce noise in the phase curve due to the sometimes large scatter in measured asteroid colours, the data can be re-extracted using a single average of the measured colours.
        However, some asteroids are known to undergo phase-reddening as they approach lower phase angles \citep[e.g.,][]{1976Icar...28...53M}, so instead we can fit a linear model to measure the reddening of the object with decreasing phase angle.
        This reddening model is then used to re-calibrate the reduced magnitudes.
        
        Taxonomic classifications of these objects are evaluated using a method described by \citet{2016P&SS..123..117P}.
        Average $G1$, $G2$ parameters for a selection of taxonomic classes compiled by \citet{2016P&SS..123..101S} are used to generate a set of one-parameter models, which can be used to obtain an estimate for the absolute magnitude $H$ even with very few data-points.
        \citet{2016P&SS..123..117P} suggest that by comparing the Bayesian Information Criterion (BIC) of each model fit to the data, a most probable classification of the asteroid can be obtained.
        To apply this technique to our parameter estimation methods, an information criterion applicable to MCMC techniques is required.
        This is selected as the Deviance Information Criterion \citep[DIC,][]{spiegelhalter2002bayesian}, an information criterion that evaluates `goodness of fit' over posterior samples and penalises model complexity.
        This statistic is calculated for each one-parameter model fit and the model with the lowest DIC indicates the asteroid classification that best supports the data.
        
\section{Target Observations and Derived Properties} \label{sec:results}
    
    \begin{deluxetable}{lrr}
        \tablecaption{Derived phase curve parameters for 8014 (1990 MF) and 19764 (2000 NF5) from PIRATE observations.\label{tab:targetParams}}
        \tablewidth{\textwidth}
        \tablehead{
            \colhead{Property} &
            \multicolumn{2}{c}{Target} \\
            \cline{2-3}
            \colhead{} &
            \colhead{8014 (1990 MF)} &
            \colhead{19764 (2000 NF5)}
        }
        \startdata
            $H_\beta$ (mag)      & $19.192 \pm 0.028$ & $16.370 \pm 0.037$ \\
            $\beta$ (mag/degree) & $0.034 \pm 0.001$  & $0.033 \pm 0.001$  \\ \hline
            $H_G$ (mag)          & $18.853 \pm 0.047$ & $16.028 \pm 0.047$ \\
            $G$                  & $0.097 \pm 0.024$  & $0.094 \pm 0.022$  \\ \hline
            $H_{G12}$ (mag)      & $18.978 \pm 0.023$ & $16.144 \pm 0.024$ \\
            $G_{12}$             & $0.656 \pm 0.063$  & $0.712 \pm 0.065$  \\ \hline
            $H_{G1,G2}$ (mag)    & $19.122 \pm 0.084$ & $16.228 \pm 0.089$ \\
            $G_1$                & $0.756 \pm 0.083$  & $0.788 \pm 0.106$  \\
            $G_2$                & $0.169 \pm 0.044$  & $0.120 \pm 0.042$  \\
        \enddata
    \end{deluxetable}

    \subsection{8014 (1990 MF)}
        
        Near-Earth Asteroid 8014 (1990 MF) was observed with PIRATE on 21 nights between 2020-05-28 and 2020-07-13, totalling 46.9 hours of observations.
        The range of phase angles was 8.4 - 91.9 degrees ($\Delta\alpha = 83.5$ degrees).
        Using the methodology described in Section~\ref{subsec:spin}, no rotational variations are detected above the noise level in the data.
        This suggests that over the viewing geometries this object was observed at, the asteroid is approximately rotationally symmetric.
        The relative light curves and corresponding scale uncertainties for this asteroid can be found in Figure~\ref{fig:8014thumbnailLCs} in the appendix.
        
\begin{figure}[t]
            \centering
            \includegraphics[width=\linewidth]{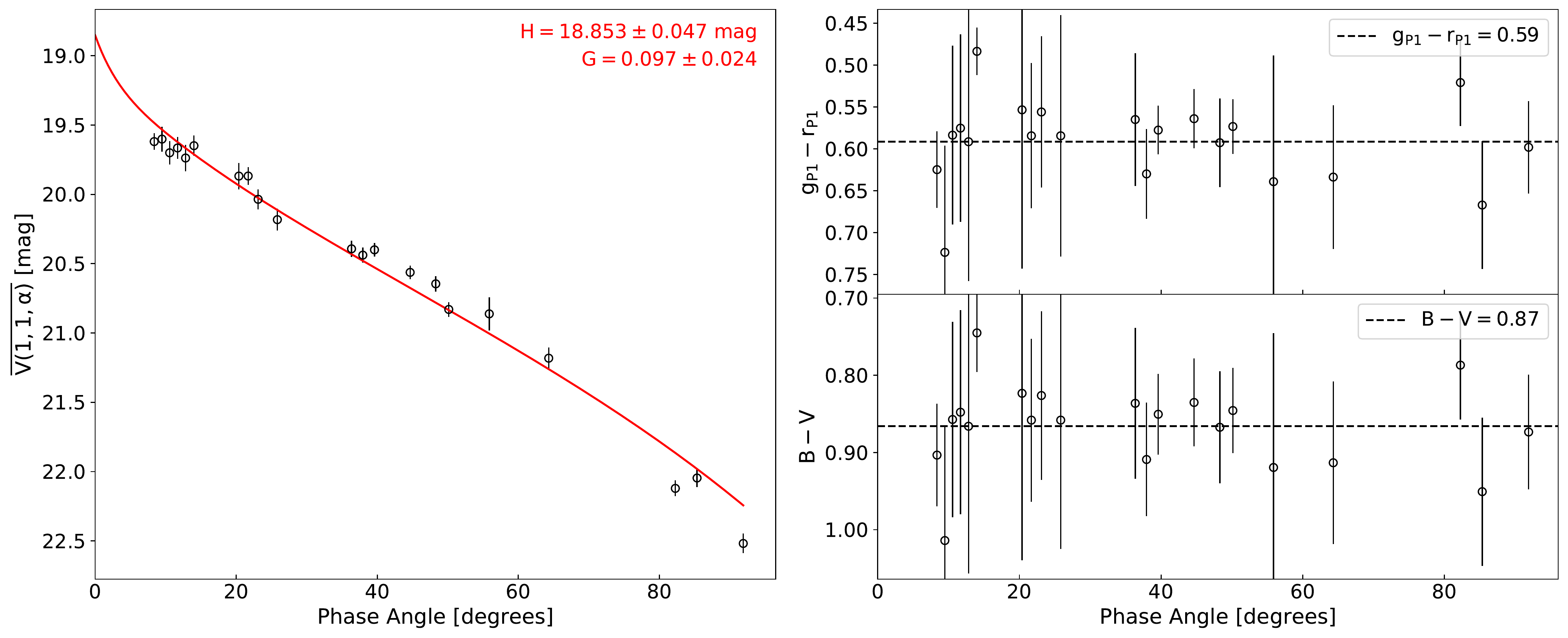}
            \caption{\textbf{Left Panel:} Phase curve data for 8014 (1990 MF). Red line represents the H, G system model for the parameters that best support the data. \textbf{Right Panel:} Measured Pan-STARRS $(g_{P1} - r_{P1})$ and calculated Johnson $(\bv)$ colour of 8014 (1990 MF) as a function of phase angle. A linear relation is fit to the data to measure the phase reddening of the asteroid. No significant phase reddening is observed for this object.}
            \label{fig:8014HG}
        \end{figure}

        The phase curve parameters that best support the data for each model are shown in Table~\ref{tab:targetParams}.
        As an example, the model using the most probable parameters in the H, G system is plotted with the phase curve data in Figure~\ref{fig:8014HG} alongside the asteroid colour variations with phase angle.
        We derive a Pan-STARRS colour $(g_{P1} - r_{P1}) = 0.59 \pm 0.01$ mag, and a $(\bv)$ colour of $0.87 \pm 0.01$ mag.
        
        The absolute magnitudes derived using these models are broadly consistent with, but not within uncertainty of, a previously derived value of 18.7 mag stated in the Minor Planet Circular (MPC 30855).
        This discrepancy is not an issue for our data since most Minor Planet Center data comes from sparse astrometric-focused observations and the accuracy of such measurements is typically much less than that of our data.
        Using the taxonomic classification method outlined previously, the classification with the slope parameters that best supports our data is a P-type model.
        However, without higher quality data close to zero phase angle this method is limited when distinguishing between the low albedo classifications (P, C, and D).
        We therefore limit our interpretation of the data to only suggest that this object may be a low albedo asteroid.
    
    \subsection{19764 (2000 NF5)}
        
\begin{figure}[t]
            \centering
            \includegraphics[width=.6\textwidth]{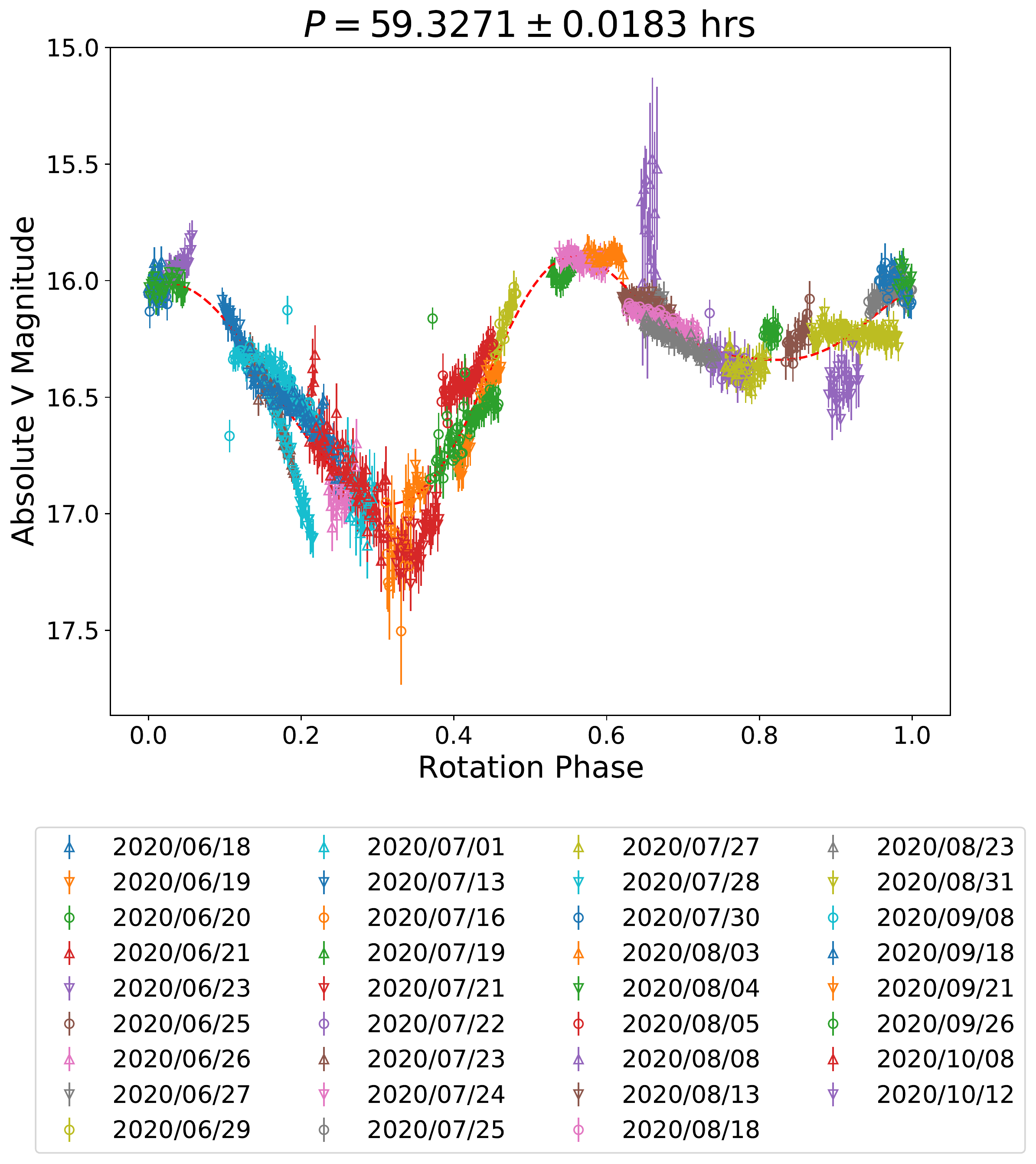}
            \caption{Best fit rotation period and 4th order Fourier series model (dotted black line) for 19764 (2000 NF5), overplotted with the phase folded reduced magnitude light curves corrected to zero phase angle (i.e the absolute magnitude) using a linear phase function. The period is calculated as $P = 59.3271\pm0.0183$ hours.}
            \label{fig:19764folded}
        \end{figure}

        Near-Earth Asteroid 19764 (2000 NF5) was observed with PIRATE on 35 nights between 2020-06-18 and 2020-10-12, totalling 117.77 hours of observations.
        The range of phase angles was 2.94 - 52.25 degrees ($\Delta\alpha = 49.31$ degrees).
        From these observations a synodic rotation period of $P = 59.3271\pm0.0183$ hours is derived (see Figure~\ref{fig:19764folded} for the phase corrected light curves folded to this rotation period).
        This is consistent with pre-published data from the Ondrejov Asteroid Photometry Project\footnote{Available at: \url{http://www.asu.cas.cz/~ppravec/newres.txt}} which suggests a rotation period of $59.3 \pm 0.1$ hours for this target.
        A maximum light curve amplitude of approximately $1.45$ mag is measured, indicating a lower bound on the axial ratios of the shape of $a/b \gtrsim 3.8$.
        The relative light curves and corresponding scale uncertainties for this asteroid can be found in Figure~\ref{fig:19764thumbnailLCs} in the appendix.
        
\begin{figure}
            \centering
            \includegraphics[width=\linewidth]{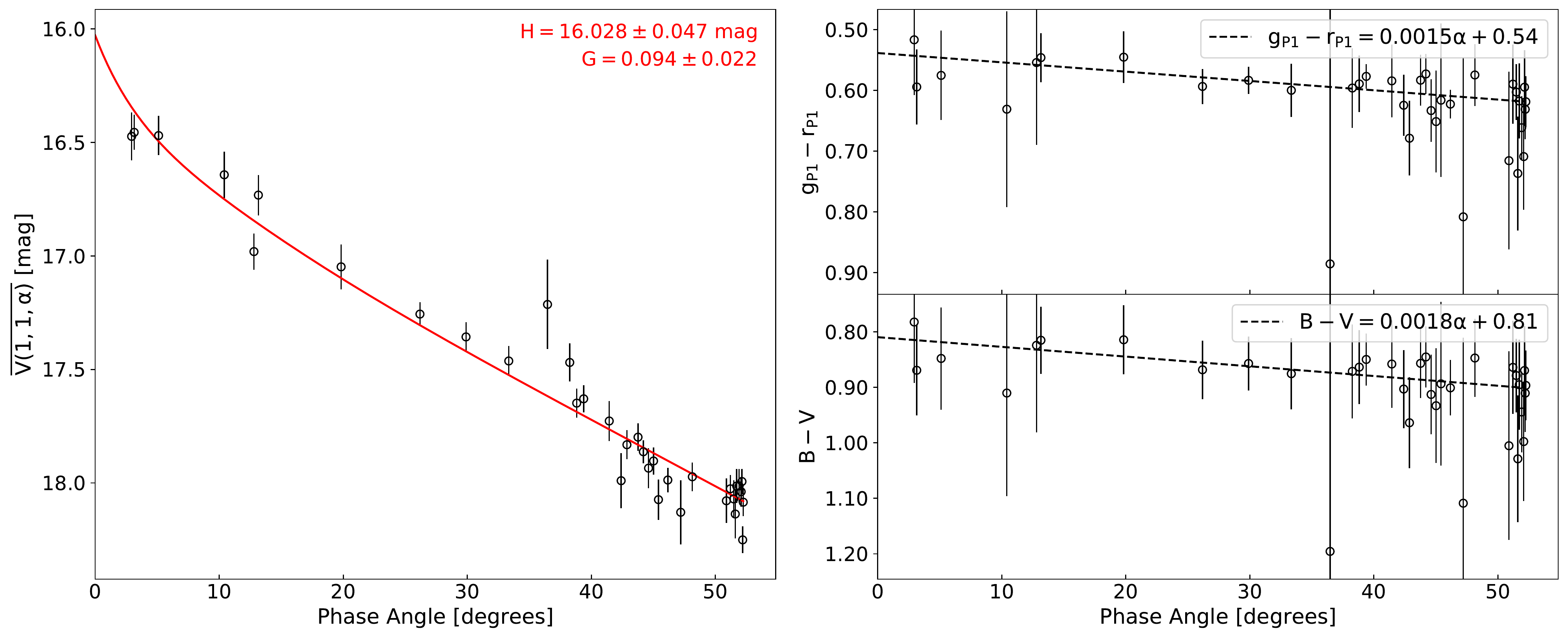}
            \caption{\textbf{Left Panel:} Phase curve data for 19764 (2000 NF5). Red line represents the H, G system model for the parameters that best support the data. \textbf{Right Panel:} Measured Pan-STARRS $(g_{P1} - r_{P1})$ and calculated Johnson $(\bv)$ colour of 19764 (2000 NF5) as a function of phase angle. A linear relation is fit to the data to measure the phase reddening of the asteroid. A very slight reddening relation is observed, but a constant colour model would also well support the data.}
            \label{fig:19764HG}
        \end{figure}

        The phase curve parameters that best support the data for each model are shown in Table~\ref{tab:targetParams}.
        The model using the most probable parameters in the H, G system is plotted over the phase curve data in Figure~\ref{fig:19764HG} alongside the asteroid colour variations with phase angle.
        The absolute magnitude and slope parameter derived using this model are not consistent with that from \citet{2012Icar..221..365P}, with this study suggesting H \& G to be $16.280 \pm 0.084$ mag \& $0.3 \pm 0.1$ respectively.
        Using the taxonomic classification method outlined previously, the classification that best supports our data is again a P-type model.
        As with 8014 (1990 MF), we limit our interpretation of the data to only suggest that this object could be a low albedo asteroid.
        However, this is inconsistent with multiple S-type spectral classifications in the literature \citep{2013Icar..225..131S,2014Icar..228..217T}.
        
        We derive a Pan-STARRS colour $(g_{P1} - r_{P1}) = 0.54 \pm 0.01$ mag with a reddening slope of $0.0015 \pm 0.0002$ mag/degree. 
        A $(\bv)$ colour of $0.834 \pm 0.021$ mag from \citet{2003Icar..163..363D} at $\alpha = 15.39$ is consistent with our calculation of $(\bv) = 0.81 \pm 0.02$ mag at zero phase angle with a reddening slope of $0.0018 \pm 0.0005$ mag/degree.        
        The observation of phase reddening is consistent with the S-type classifications from the literature, as material with an Olivine spectral feature (e.g. S-type and Q-type asteroids) is expected to undergo more spectral reddening than other taxonomic classes that do not have this spectral feature \citep{2012Icar..220...36S}.
        This `misclassification' and the inconsistency of our results with previously published data may indicate that there are modulations to the phase curve between or during apparitions that are dependent on shape and viewing geometry \citep{Rozitis2020influence}.
        With our derived light curve amplitude suggesting an elongated object, these modulations due to changing aspect may be significant between apparitions.
        The deviations in light curve shape apparent over the single apparition observed (Figure~\ref{fig:19764folded}) also provide evidence for the effect of changing aspect.
        This effect of shape and aspect on asteroid phase curves will be addressed in a future publication.

\section{Observatory and Pipeline Performance} \label{sec:observatory}
    
\begin{figure}[t]
        \centering
        \includegraphics[width=\linewidth]{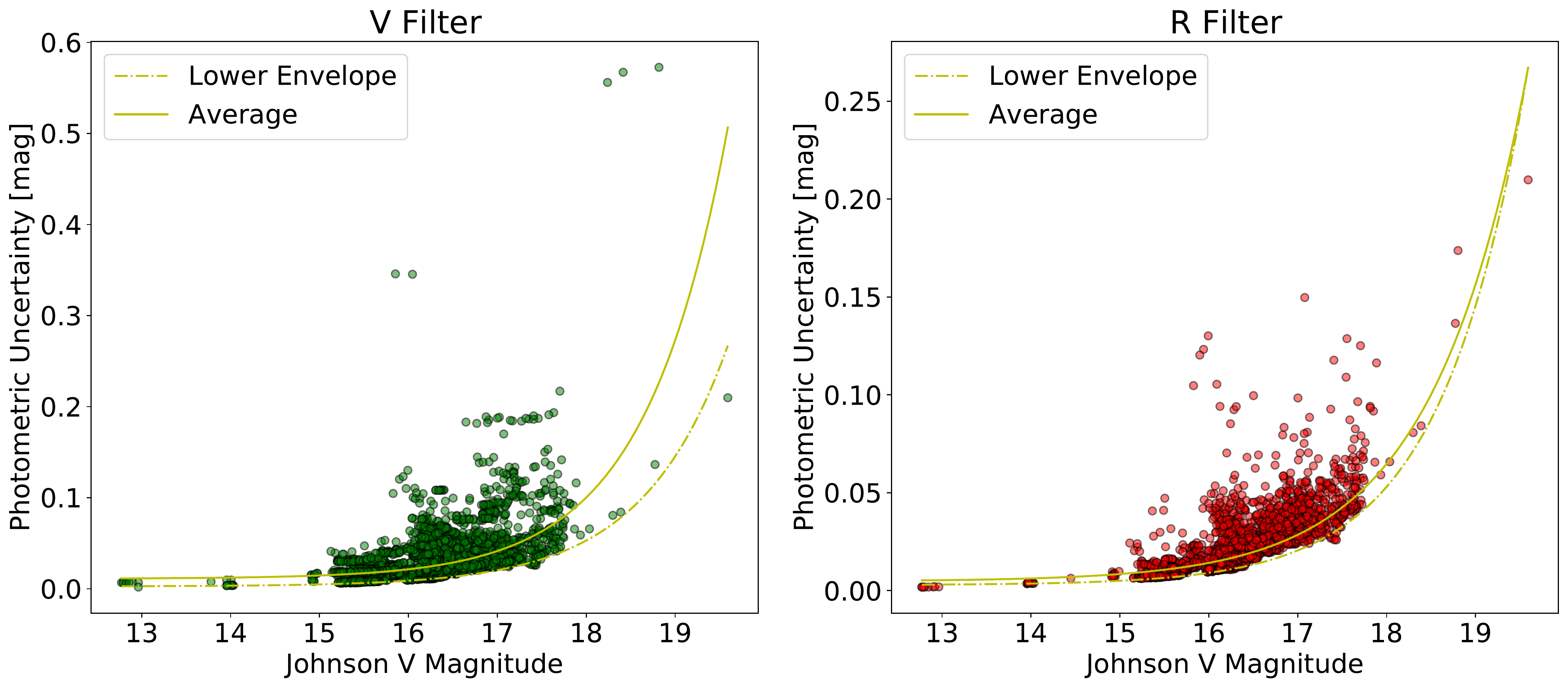}
        \caption{Measured photometric uncertainty using the V filter (left panel) and R filter (right panel) plotted against the calculated Johnson V magnitude of the target. Photometric uncertainty is observed to decrease exponentially with brighter objects as expected. Uncertainties are generally lower in the R filter than in the V filter due to the typically red asteroid colours and higher quantum efficiency of the CCD at red wavelengths.}
        \label{fig:photometricPerformance}
    \end{figure}

    A primary goal when observing faint objects with small aperture telescopes is to improve photometric performance.
    The photometric performance of PIRATE is demonstrated by plotting the derived photometric uncertainties as a function of Johnson V magnitude for over 5500 individual measurements of 15 different asteroids over a 10 month period.
    Images taken with the Johnson R filter generally provide a lower photometric uncertainty  (Figure~\ref{fig:photometricPerformance}, right panel) than images taken with the Johnson V filter  (Figure~\ref{fig:photometricPerformance}, left panel).
    This is expected as asteroid spectra typically peak towards `redder' wavelengths and the CCD quantum efficiency is higher at redder wavelengths.
    Table~\ref{tab:photPrecision} lists the average and best case photometric uncertainties for each filter, for a range of Johnson V magnitudes.
    The photometric uncertainty begins to become dominant compared to the typical scale uncertainty ($0.049\pm0.008$ mag) at $17.66$ mag for the R filter and $17.18$ mag for the V filter.
    This marker indicates a point where uncertainty in the period and Fourier components will begin to rise rapidly for objects with low amplitude rotation, although useful data will still be obtained for highly asymmetric objects with high amplitude light curves.
    
    \begin{deluxetable}{ccccc}
        \tablecaption{Average and best case photometric uncertainties for objects observed in the V and R filters on PIRATE, for a range of Johnson V magnitudes.\label{tab:photPrecision}}
        \tablewidth{\textwidth}
        \tablehead{
            \colhead{Johnson V} &
            \multicolumn{2}{c}{Average Uncertainty} & 
            \multicolumn{2}{c}{Best Case Uncertainty} \\ 
            \colhead{(mag)} &
            \multicolumn{2}{c}{(mag)} &
            \multicolumn{2}{c}{(mag)} \\ \cline{2-5}
            \colhead{} &
            \colhead{V filter} &
            \colhead{R filter} &
            \colhead{V filter} &
            \colhead{R filter}
        }
        \startdata
            13 & 0.012 & 0.005 & 0.003 & 0.003 \\
            14 & 0.012 & 0.006 & 0.003 & 0.004 \\
            15 & 0.015 & 0.008 & 0.005 & 0.005 \\
            16 & 0.021 & 0.014 & 0.009 & 0.009 \\
            17 & 0.041 & 0.029 & 0.020 & 0.020 \\
            18 & 0.100 & 0.065 & 0.053 & 0.053 \\
            19 & 0.273 & 0.156 & 0.145 & 0.145 \\
        \enddata
    \end{deluxetable}
    
\begin{figure}[t]
        \centering
        \includegraphics[width=.6\linewidth]{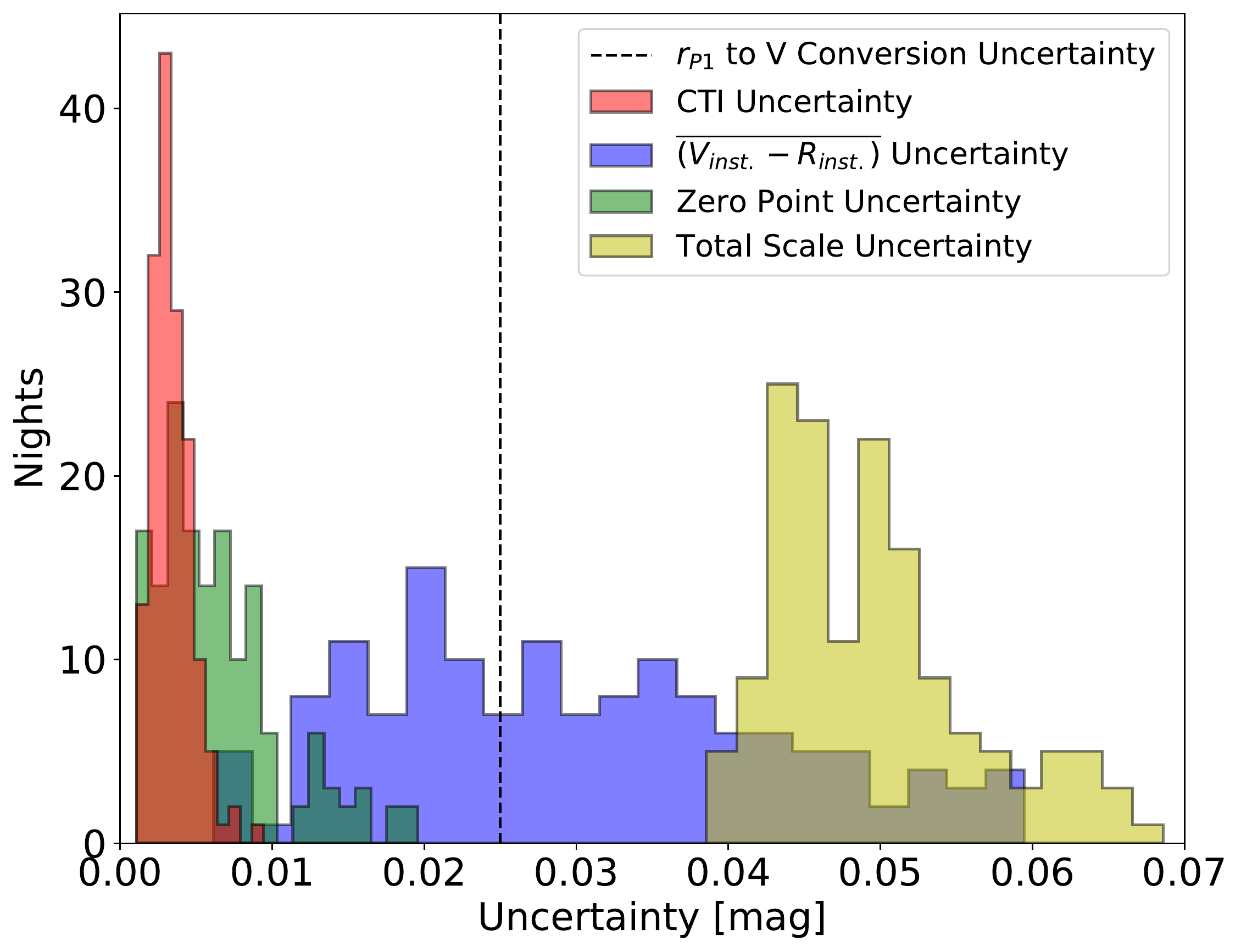}
        \caption{Distributions of colour transformation intercept uncertainties, instrumental colour uncertainties, nightly zero point uncertainties, and the total scale uncertainties. Distributions are limited to observations where scale uncertainty is less than 0.07 mag to avoid sporadic tail values concealing the typical values in the histograms.}
        \label{fig:scaleUnc}
    \end{figure}

    The average scale uncertainty when calibrating PIRATE observations is approximately $0.049\pm0.006$ mag (see Figure~\ref{fig:scaleUnc}).
    This scale uncertainty is a combination of the uncertainty on: the colour transformation coefficients, the colour term, the nightly zero points, the instrumental colour, and the photometric system conversions.
    The colour transformation gradient and the colour term have comparably low uncertainties to the overall scale uncertainty due to their derivation from a large ensemble of observations.
    Only the colour transformation intercept, instrumental colour, and zero point are measured each night.
    The uncertainty distributions of these parameters plus the total scale uncertainty are shown in Figure~\ref{fig:scaleUnc}.
    The mean instrumental colour uncertainty is $0.033\,$mag, the mean CTI uncertainty is $0.003\,$mag, and the mean zero point uncertainty is $0.006\,$mag.
    The uncertainty in the linear photometric transformation equation (Equation~\ref{eq:rP1toV}) between $r_{P1}$ and Johnson V is 0.025 mag \citep{2012ApJ...750...99T}, and will remain constant unless updated photometric conversions are calculated.
    The instrumental colour uncertainty is therefore the most significant contributor to the scale uncertainty out of the four major contributors for the majority of objects.
    The uncertainty in the instrumental colour measurement (and subsequently the total scale uncertainty) can be improved by observing brighter asteroids.
    
\begin{figure}[t]
        \centering
        \subfloat[]{\includegraphics[width=0.49\textwidth]{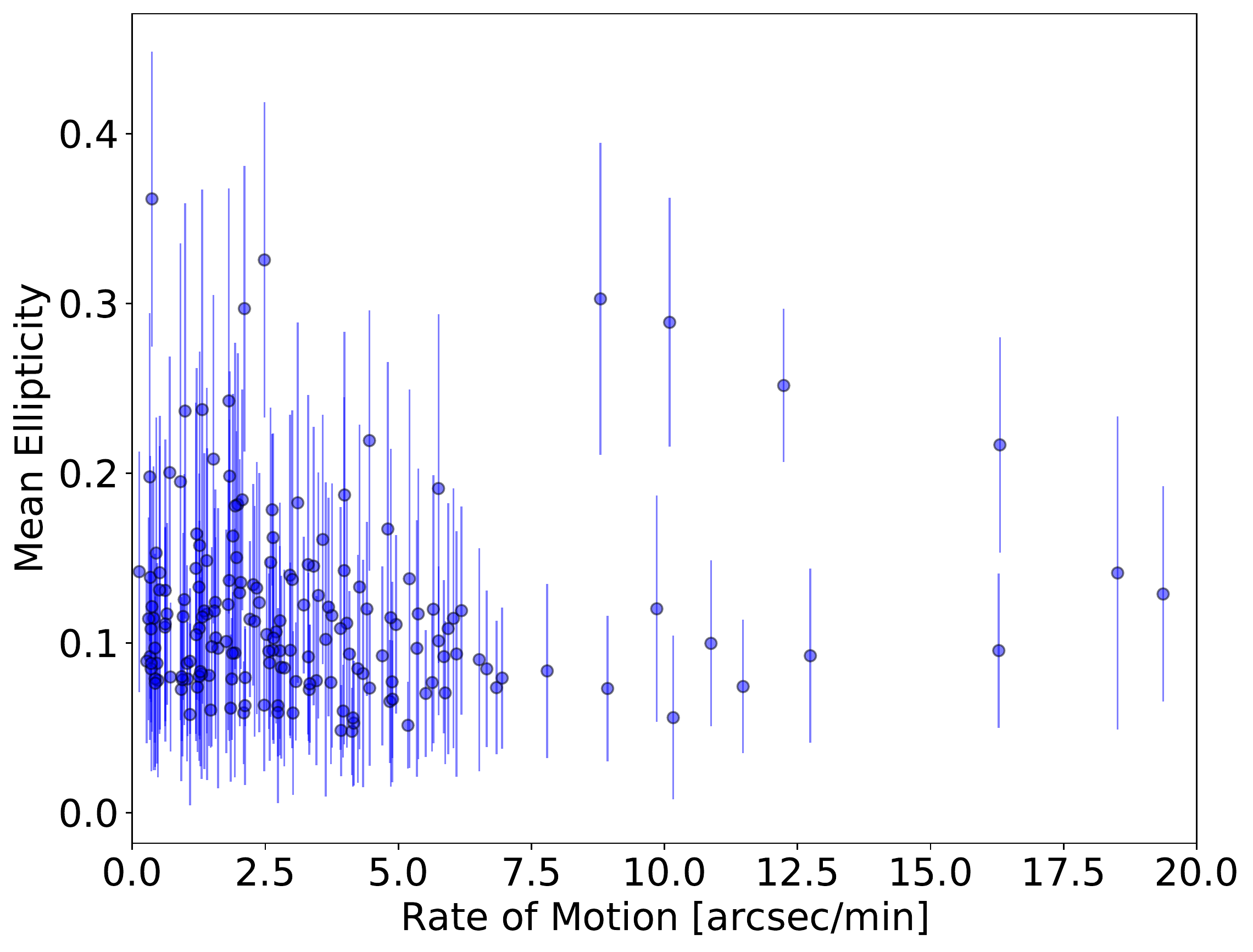}}
        \subfloat[]{\includegraphics[width=0.49\textwidth]{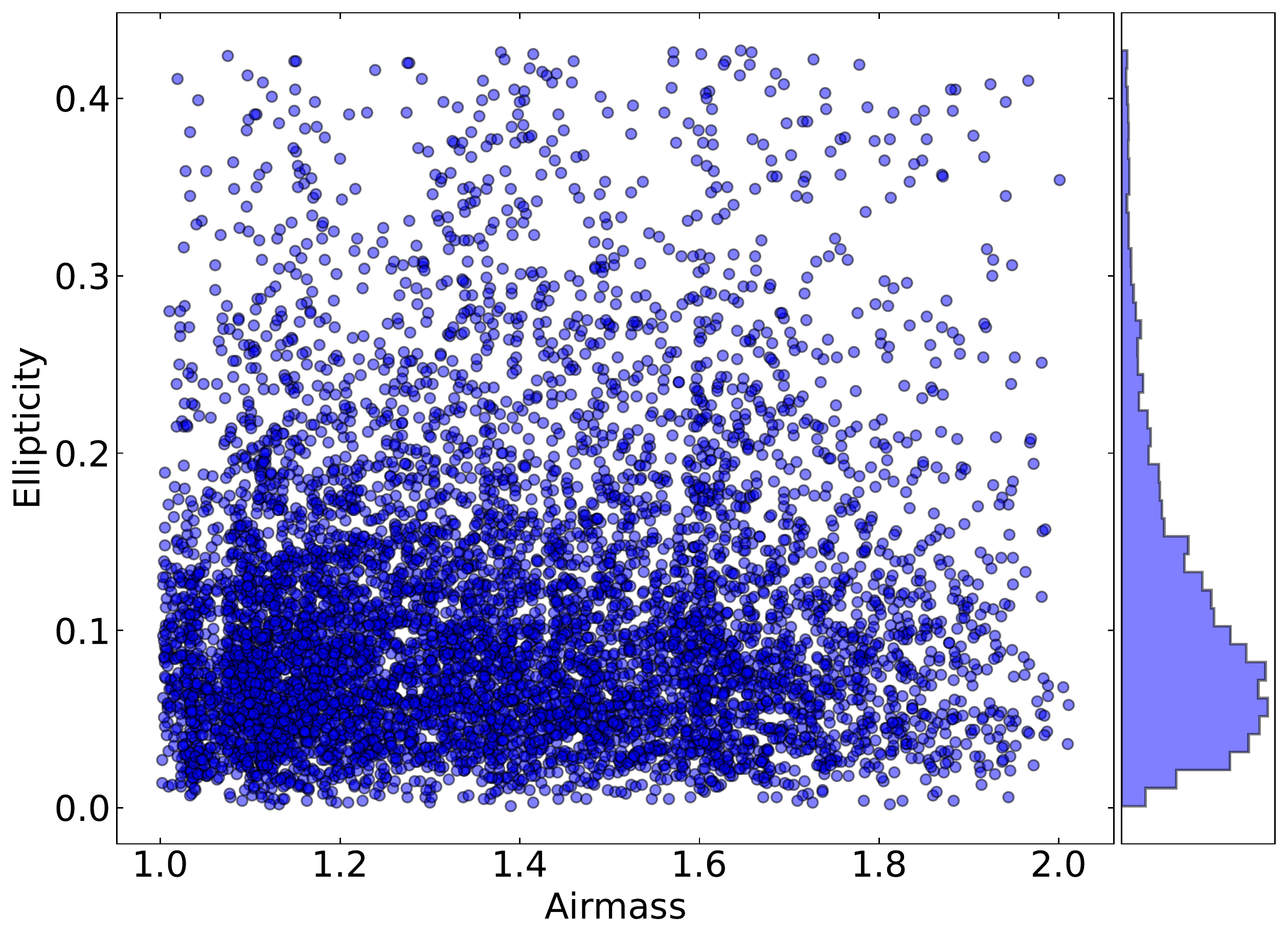}}
        \caption{\textbf{(a)} Average ellipticity on a given night in images of tracked objects, plotted against the average rate of motion during the observing run. Uncertainties on the ellipticities represent the scatter in ellipticities during the observing run, calculated as the standard deviation. \textbf{(b)} All ellipticity measurements in images of tracked objects over the observing campaign, plotted against airmass. The median ellipticity over the observing campaign is 0.085.}
        \label{fig:trackingPerformance}
    \end{figure}

    It is also important to characterise the performance of the differential tracking by the mount, as accumulating large tracking errors will impact the uncertainty in our data.
    A poorly tracked asteroid is expected to have a larger ellipticity in an image than that of a well tracked asteroid.
    Figure~\ref{fig:trackingPerformance}(a) shows the average ellipticity of various targets over different nights plotted against the target rates of motion, with uncertainties representing the scatter in individual measurements throughout the night.
    It was expected that some degradation in tracking performance would be observed with increased asteroid on-sky rate of motion, although such a trend is not observed for these observations.
    This result is biased by a significant absence of large rates of motion in the data due to selection criteria for the targets.
    However, even among the data at smaller rates of motion no trend is observed beyond the scatter of the data.
    This is a sign that the telescope mount is able to handle a suitable range of tracking speeds for near-Earth Asteroid observations without degradation of image quality.
    Degradation in tracking performance may still arise at even faster tracking speeds, but due to the limitations of the data extraction process it is not likely that targets will be observed over this range.
    
    No trend in ellipticity is observed with airmass (Figure~\ref{fig:trackingPerformance}(b)), showing that tracking performance or image quality does not significantly vary with on-sky position.
    The source of the scatter in ellipticities within each night is assumed to be random scatter as no variables are identified to trend with this variation.
    The median ellipticity of all observations is $0.085$, indicating good tracking performance overall by the mount.
    Some level of tracking error will always be expected due to the usage of average tracking rates as discussed in Section~\ref{subsec:observationStrategy}.

\newpage
\section{Summary} \label{sec:summary}
    
    In order to maximise the scientific output of the PIRATE facility, the existing standard data reduction processes have been optimised to minimise noise sources where possible.
    Bias correction is now conducted using a bias structure map that is scaled according to a mean bias level relationship with dome temperature.
    Dark correction is no longer performed due to the extremely low dark current present in the CCD at operating temperatures, and due to various scaling issues present with a sample of pixels.
    The variability of the flatfield frames has been characterised to optimise the timescale over which flatfield frames can be combined to produce the best possible master flatfield each night.
    These procedures and investigations are applicable to other small-aperture facilities, and in particular the flatfield optimisation may help to optimise their scientific output in the context of asteroid science.
    
    We have presented the design and operation of a new precise photometric calibration pipeline for targeted asteroid observations with PIRATE.
    For a 17th magnitude (Johnson V) asteroid observed in the instrumental R filter, the photometric uncertainty can be expected to be approximately $0.029\,$mag on average, and $0.020\,$mag in the best case scenario.
    This comfortably allows for the rotational characterisation of asteroids with light curve amplitudes $\gtrapprox 0.1$ mag, with even lower rotational amplitudes resolvable as brightness increases.
    The methods outlined provide an average calibration uncertainty of $0.049\,$mag, allowing for accurate photometric calibration and subsequent phase curve extraction of NEAs with a small telescope.
    The comparable photometric and calibration uncertainties at $\sim17.16$ mag in the V filter allow us to set this as a marker below which we prefer to observe objects with high amplitude light curves where possible.
    The observatory is not found to suffer an increase in tracking errors with faster target rates of motion (up to the limits observed during this work), nor with increased airmass, indicating good tracking performance of the mount.
    
    The capabilities of the hardware and software are demonstrated through the photometric characterisation of near-Earth Asteroids 8014 (1990 MF) and 19764 (2000 NF5).
    8014 (1990 MF) is found to be roughly rotationally symmetric at this aspect and no rotation period is identified for this target.
    Phase curve parameters are extracted for this object in four photometric systems.
    Using the one-parameter phase curve fitting method described in \citet{2016P&SS..123..117P} this object is deemed likely to be a low albedo asteroid.
    19764 (1990 MF) is calculated to have a $59.3271\pm0.0183$ hour rotation period, comparable to prior estimates from unpublished data.
    This asteroid is likely an elongated object due to its observed 1.45 magnitude light curve amplitude.
    This object is also deemed likely to have a low albedo from the phase curve data.
    Our results are inconsistent with previous phase curve and taxonomic studies of this asteroid, and this may be indicative of shape dependent modulations to the phase curve over different apparitions \citep{Rozitis2020influence}, which may be significant for NEAs.
    An evaluation of the potential errors introduced into asteroid phase curves by shape and aspect effects, as well as identification of objects that have these effects present in their phase curves, will be the subject of future work.
    
    The high quality of the calibration and extracted phase curves demonstrates the important contribution that small aperture facilities can make to the field of asteroid science beyond rotation period estimation.
    Observers with a comparable telescope capable of differential tracking are shown to be able to derive physical properties of asteroids using targeted observing campaigns and optimised data collection, reduction, and extraction techniques.

\newpage
\begin{acknowledgments}

This research was made possible through the OpenSTEM Labs, an initiative funded by HEFCE and by the Wolfson Foundation.
S.~L. Jackson is funded by the Science and Technology Facilities Council under grant ST/T506321/1 (project reference: 2284918).
U.~C. Kolb, and S.~F. Green are funded by the Science and Technology Facilities Council under grants ST/T000295/1 and ST/T000228/1 respectively.
The authors thank Sybilla Technologies, Baader Planetarium, and the Instituto de Astrof\'isica de Canarias (IAC) for their assistance developing new capabilities and supporting the continuing operation of the OpenScience Observatories.
The authors thank the anonymous reviewer for their helpful comments on the initial version of the manuscript.
Data presented in this work may be accessed upon reasonable request to the corresponding author.

\end{acknowledgments}

\software{
    This work has made use of the following software;
    \textsc{Abot}\textsuperscript{\texttrademark} \citep{2014SPIE.9152E..1CS},
    \textsc{Astrometry.net} \citep{2010AJ....139.1782L},
    \textsc{Astropy} \citep{2013A&A...558A..33A},
    \textsc{Astroquery} \citep{2019AJ....157...98G},
    \textsc{convexinv} \citep{2001Icar..153...37K,2010A&A...513A..46D},
    \textsc{Emcee} \citep{2013PASP..125..306F},
    \textsc{JPL Horizons} \citep{2015IAUGA..2256293G},
    \textsc{Matplotlib} \citep{2005ASPC..347...91B},
    \textsc{NumPy} \citep{2020Natur.585..357H},
    \textsc{Pandas} \citep{jeff_reback_2021_4572994},
    \textsc{SciPy} \citep{2020NatMe..17..261V},
    \textsc{Source Extractor} \citep{1996A&AS..117..393B}
}

\bibliographystyle{aasjournal}
\bibliography{references}{}

\appendix

\section{Individual Light Curve Plots}
    \begin{figure}[H]
        \centering
        \includegraphics[width=.8\textwidth]{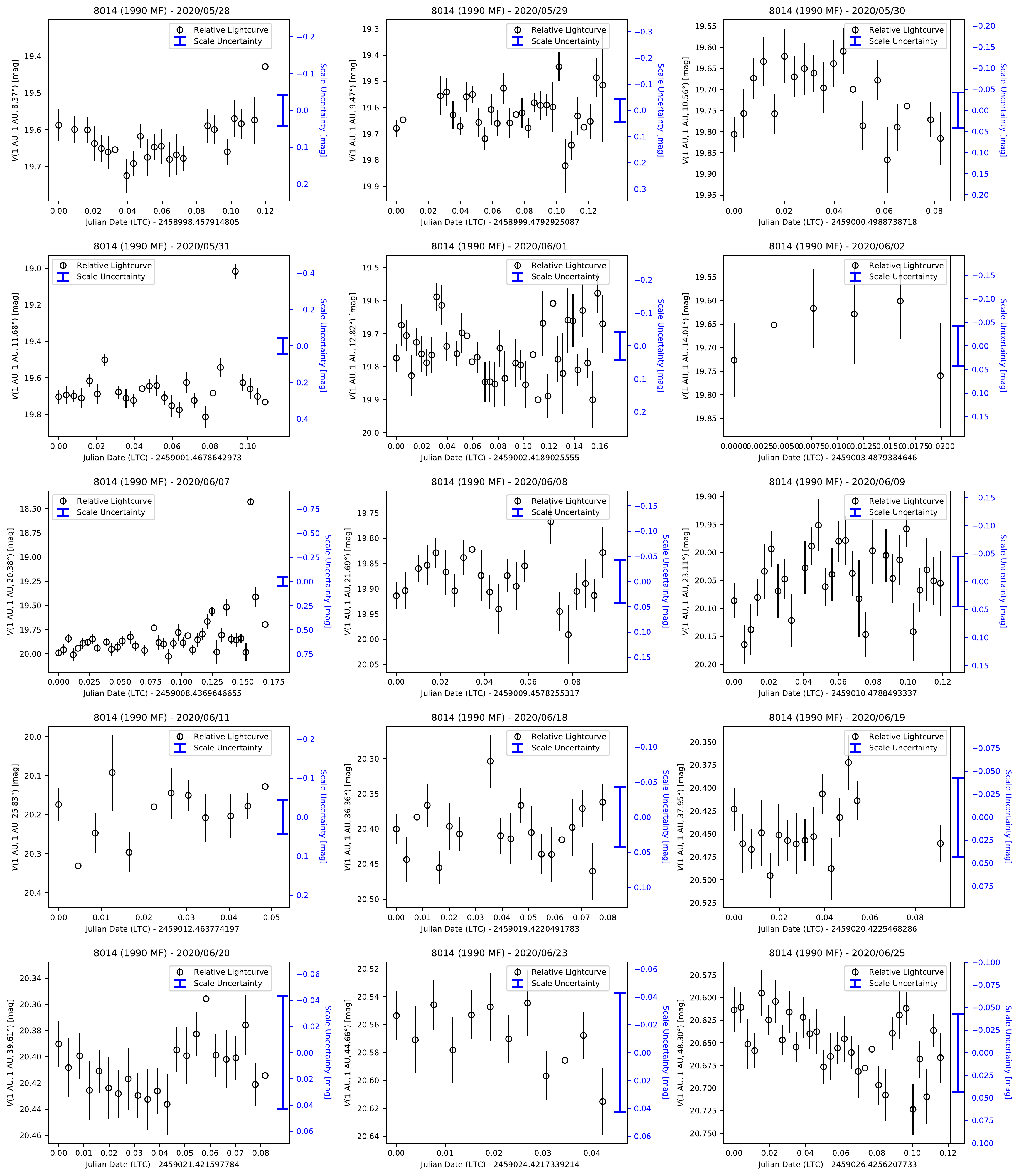}
        \caption{Individual lightcurves of 8014 (1990 MF), plotted in days since start of observations that night. Uncertainties on each data point are the relative magnitude uncertainties, the blue error bars are the scale uncertainties calculated for each night.}
        \label{fig:8014thumbnailLCs}
    \end{figure}
    \begin{figure}[t]
        \ContinuedFloat
        \centering
        \includegraphics[width=.9\textwidth]{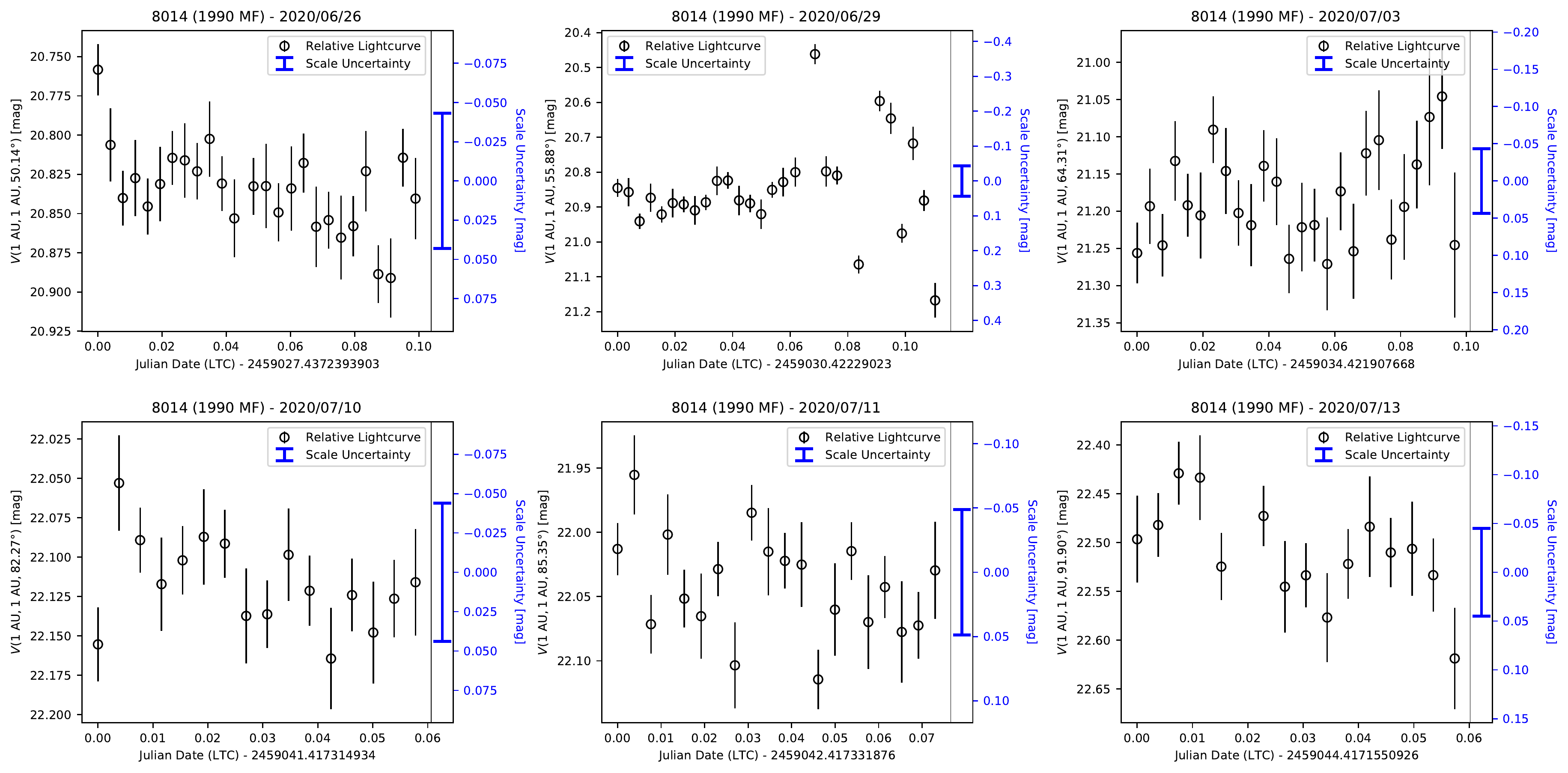}
        \caption{\textbf{(cont.)}}
    \end{figure}
    \begin{figure}[H]
        \centering
        \includegraphics[width=.9\textwidth]{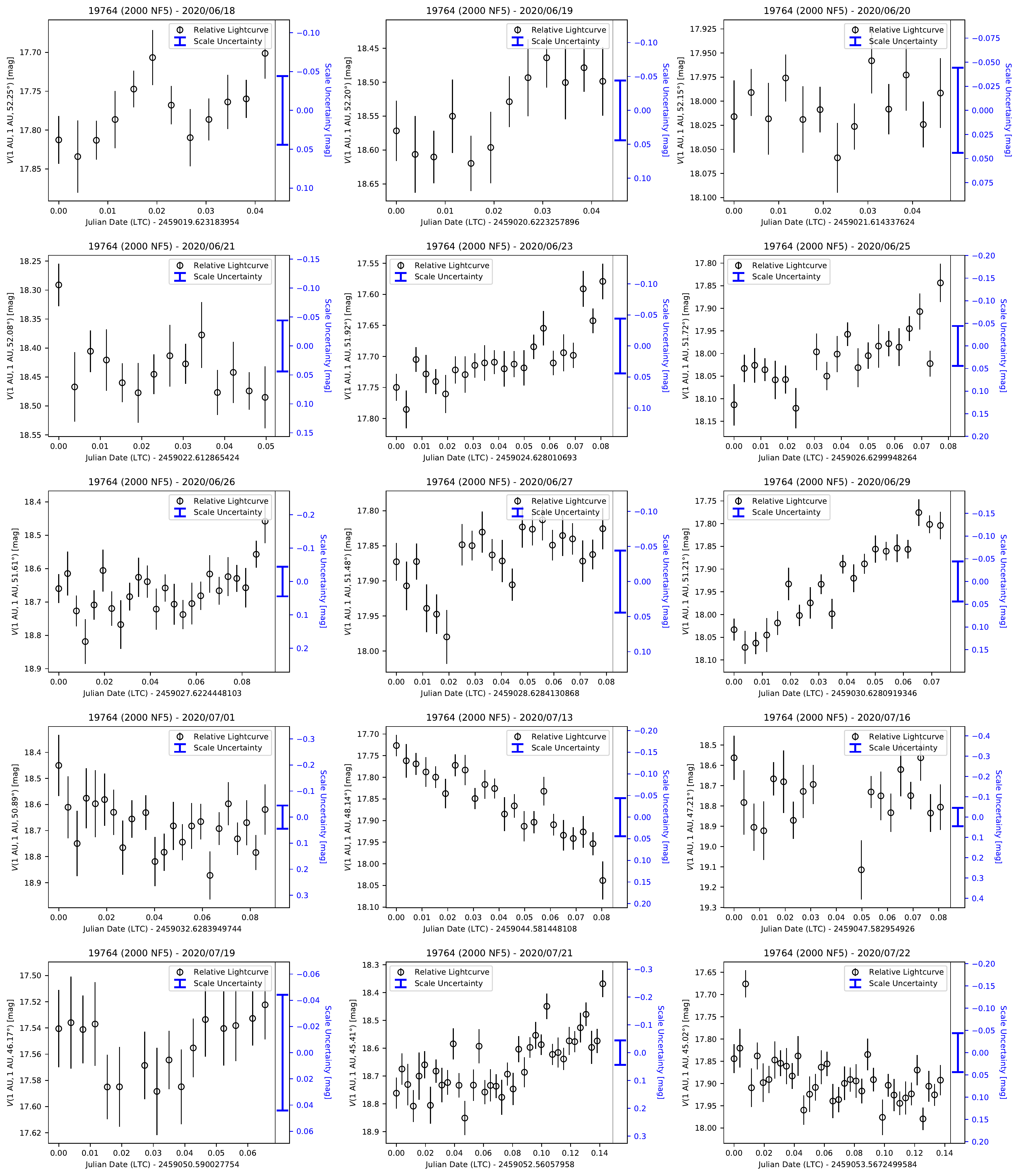}
        \caption{Individual lightcurves of 19764 (2000 NF5), plotted in days since start of observations that night. Uncertainties on each data point are the relative magnitude uncertainties, the blue error bars are the scale uncertainties calculated for each night.}
        \label{fig:19764thumbnailLCs}
    \end{figure}
    \begin{figure}[t]
        \ContinuedFloat
        \centering
        \includegraphics[width=.9\textwidth]{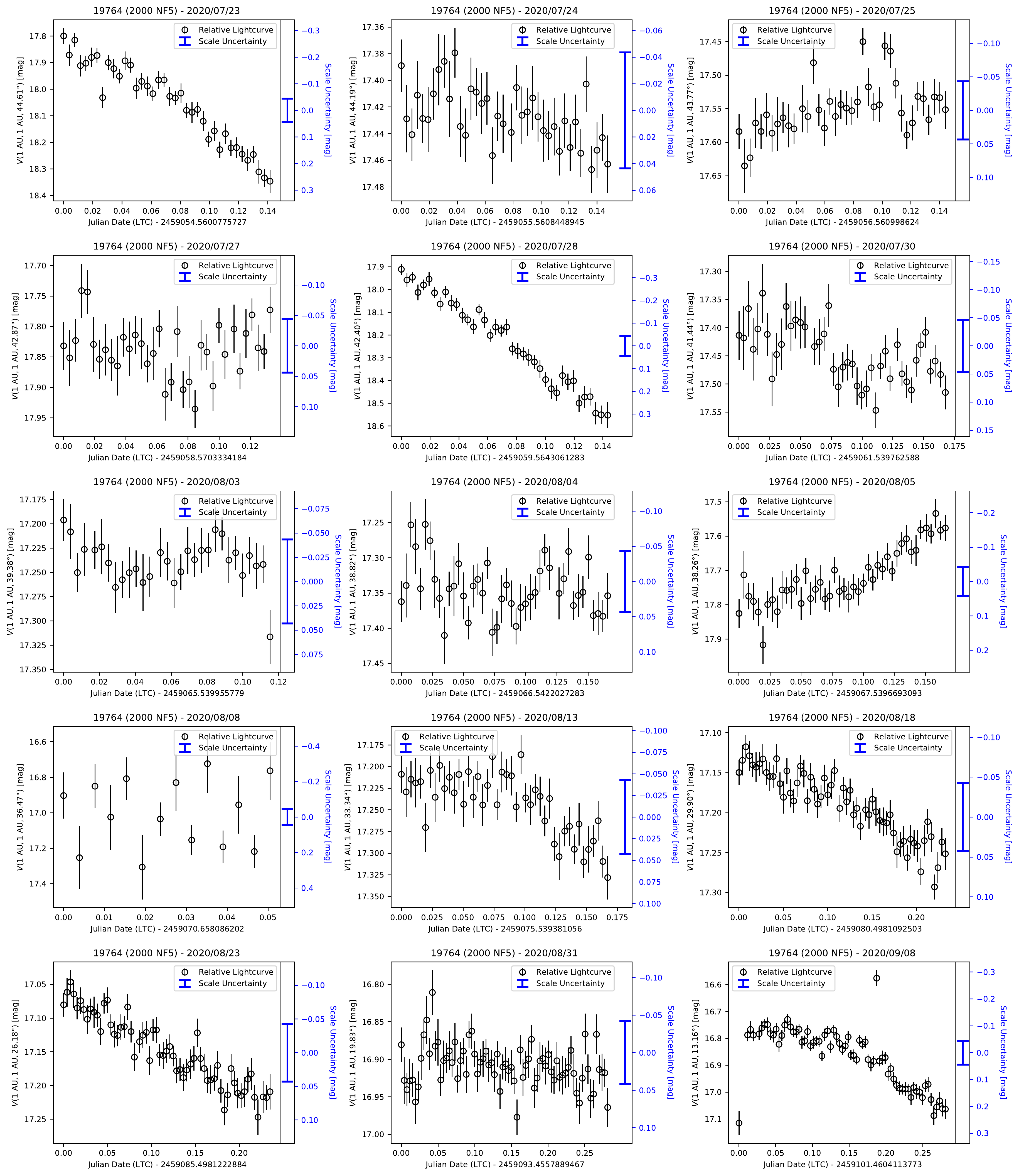}
        \caption{\textbf{(cont.)}}
    \end{figure}
    \begin{figure}[t]
        \ContinuedFloat
        \centering
        \includegraphics[width=.9\textwidth]{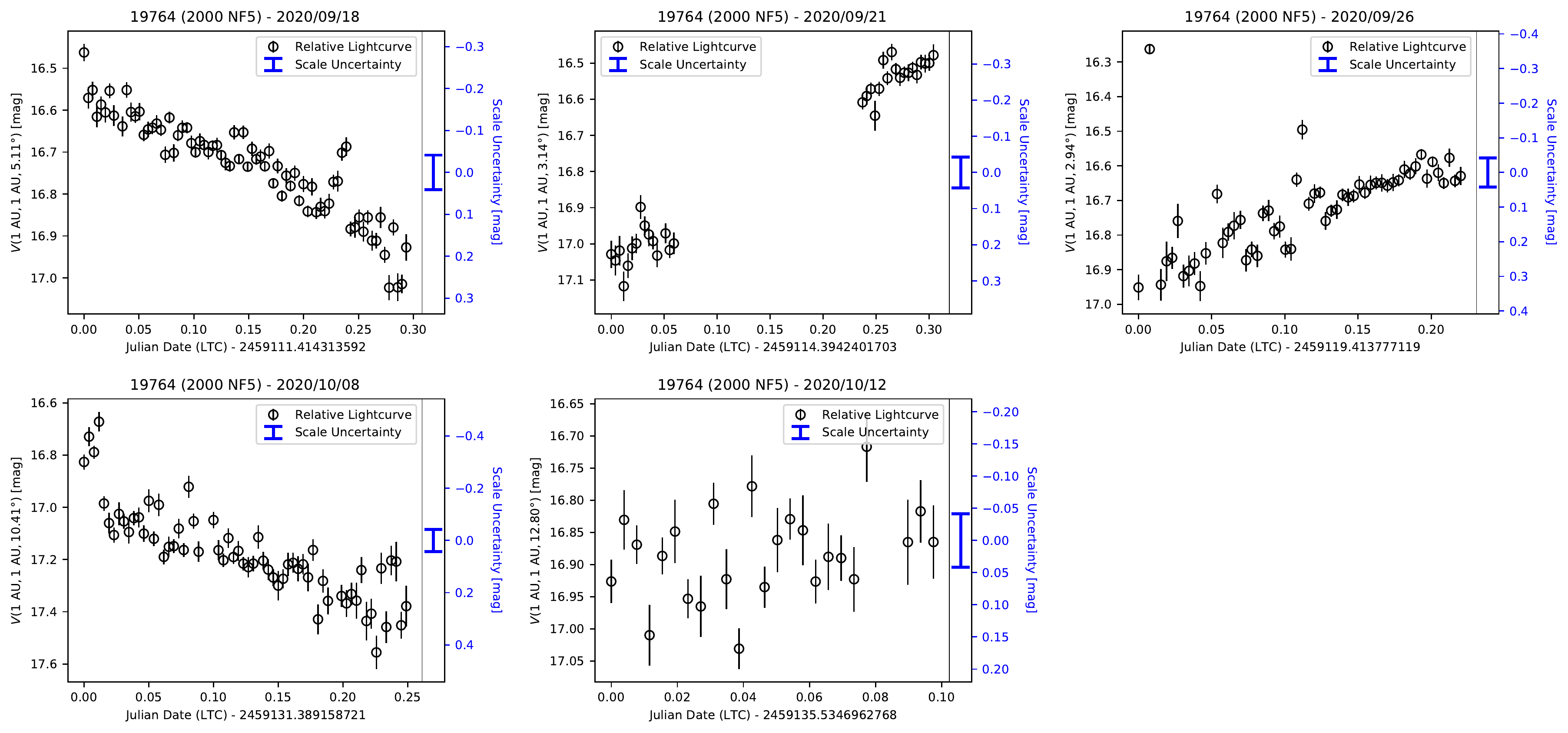}
        \caption{\textbf{(cont.)}}
    \end{figure}
    
\end{document}